\documentclass[twocolumn,usenatbib]{aastex62}

\usepackage{amssymb, amsmath}

\graphicspath{{./}{Figures/}}

\shorttitle{A New Class of X-Ray Tails}
\shortauthors{Sheardown et al.}
\begin{document}

\title{A New Class of X-Ray Tails of Early-Type Galaxies and Subclusters in Galaxy Clusters - Slingshot Tails vs Ram Pressure Stripped Tails}

\correspondingauthor{Alex Sheardown, Elke Roediger}
\email{A.Sheardown@2011.hull.ac.uk, E.Roediger@hull.ac.uk}

\author[0000-0002-0543-7143]{Alex Sheardown}
\affiliation{E.A. Milne Centre for Astrophysics,
Department of Physics and Mathematics, University of Hull,
Hull, HU6 7RX, UK}

\author[0000-0001-7125-4137]{Thomas M. Fish}
\altaffiliation{Visiting Scientist, SAO}
\affiliation{E.A. Milne Centre for Astrophysics,
Department of Physics and Mathematics, University of Hull,
Hull, HU6 7RX, UK}
\affiliation{Harvard-Smithsonian Center for Astrophysics, 60 Garden Street,
Cambridge, MA 02138, USA}

\author[0000-0003-2076-6065]{Elke Roediger}
\affiliation{E.A. Milne Centre for Astrophysics,
Department of Physics and Mathematics, University of Hull,
Hull, HU6 7RX, UK}

\author{Matthew Hunt}
\affiliation{E.A. Milne Centre for Astrophysics,
Department of Physics and Mathematics, University of Hull,
Hull, HU6 7RX, UK}

\author[0000-0003-3175-2347]{John ZuHone}
\affiliation{Harvard-Smithsonian Center for Astrophysics, 60 Garden Street,
Cambridge, MA 02138, USA}

\author{Yuanyuan Su}
\affiliation{Harvard-Smithsonian Center for Astrophysics, 60 Garden Street,
Cambridge, MA 02138, USA}

\author[0000-0002-0765-0511]{Ralph P. Kraft}
\affiliation{Harvard-Smithsonian Center for Astrophysics, 60 Garden Street,
Cambridge, MA 02138, USA}

\author[0000-0003-0297-4493]{Paul Nulsen}
\affiliation{Harvard-Smithsonian Center for Astrophysics, 60 Garden Street,
Cambridge, MA 02138, USA}
\affiliation{ICRAR, University of Western
Australia, 35 Stirling Hwy, Crawley, WA 6009, Australia}

\author[0000-0002-0322-884X]{Eugene Churazov}
\affiliation{Max-Planck-Institut f{\"u}r  Astrophysik, Karl-Schwarzschild-Stra{\ss}e 1,
D-85741 Garching, Germany}

\author[0000-0002-9478-1682]{William Forman}
\affiliation{Harvard-Smithsonian Center for Astrophysics, 60 Garden Street,
Cambridge, MA 02138, USA}

\author{Christine Jones}
\affiliation{Harvard-Smithsonian Center for Astrophysics, 60 Garden Street,
Cambridge, MA 02138, USA}

\author[0000-0003-4917-7803]{Natalya Lyskova}
\affiliation{National Research University Higher School of Economics, 
Myasnitskaya str. 20, Moscow 101000, Russia}
\affiliation{Space Research Institute (IKI), 
Profsoyuznaya 84/32, Moscow 117997, Russia}

\author[0000-0001-7917-3892]{Dominique Eckert}
\affiliation{Department of Astronomy, University of Geneva, ch. d’Ecogia 16, 1290 Versoix, Switzerland}

\author[0000-0002-6385-8501]{Sabrina De Grandi}
\affiliation{INAF-Osservatorio Astronomico di Brera, via E. Bianchi 46, 23807 Merate, Italy}



\begin{abstract}
We show that there is a new class of gas tails - slingshot tails - which form as a subhalo (i.e. a subcluster or early-type cluster galaxy) moves away from the cluster center towards the apocenter of its orbit. These tails can point perpendicular or even opposite to the subhalo direction of motion, not tracing the recent orbital path.  Thus, the observed tail direction can be misleading, and we caution against naive conclusions regarding the subhalo's direction of motion based on the tail direction. A head-tail morphology of a galaxy's or subcluster's gaseous atmosphere is usually attributed to ram pressure stripping and the widely applied conclusion is that gas stripped tail traces the most recent orbit. However, during the slingshot tail stage, the subhalo is not being ram pressure stripped (RPS) and the tail is shaped by tidal forces more than just the ram pressure. Thus, applying a classic RPS scenario to a slingshot tail leads not only to an incorrect conclusion regarding the direction of motion, but also to incorrect conclusions in regard to the subhalo velocity, expected locations of shear flows, instabilities and mixing. We describe the genesis and morphology of slingshot tails using data from binary cluster merger simulations, discuss their observable features and how to distinguish them from classic RPS tails. We identify three examples from the literature that are not RPS tails but slingshot tails and discuss other potential candidates.
\end{abstract}

\keywords{galaxies: clusters: general --- galaxies: clusters: intracluster medium --- X-rays: galaxies --- X-rays: galaxies: clusters --- methods: numerical}


 \section{Introduction} \label{sec:intro}
Galaxy clusters grow through the sequential merging and accretion of galaxies, groups and subclusters (\citealt{Kravtsov2012b}). As one of the latter begins the merging process, it must traverse the intra-cluster medium (ICM) of its host cluster. This motion through the ICM acts as a head wind on a galaxy or subcluster, producing a ram pressure which progressively strips its gaseous atmosphere (\citealt{Gunn1972}, \citealt{Nulsen1982a}). This stripped gaseous atmosphere appears as an X-ray bright tail downstream, producing a head-tail structure and has been used to account for many observed objects e.g. in the Virgo cluster (M86: \citealt{Forman1979}; \citealt{Randall2008a}, M49: \citealt{Irwin1996}; \citealt{Kraft2011a} , M89: \citealt{Machacek2006a}; M60: \citealt{Randall2004}, \citealt{Wood2017}), NGC 4839 in Coma (\citealt{Neumann2003}, Lyskova et al. submitted), and NGC 1404 in Fornax (\citealt{Jones1997a}, \citealt{Machacek2005a}; \citealt{Su2017e}). In recent years, new X-ray tails have been discovered in several clusters at larger cluster-centric radii, e.g., in Hydra A (\citealt{DeGrandi2016b}), Abell 2142 (\citealt{Eckert2017}), and Abell 85 (\citealt{Ichinohe2015a}).\par

The gas stripping of a subcluster or early-type galaxy in the ICM of a larger, more massive cluster (the primary cluster) is very much the same process. Many simulations have confirmed that ram pressure stripping of the secondary potential is a viable process (\citealt{Gisler1976}; \citealt{Takeda1984a}; \citealt{Stevens1999a}; \citealt{Toniazzo2001}; \citealt{Acreman2003}; \citealt{McCarthy2008}, among others), and produces the expected downstream gas tail. However, gas stripping of the secondary's atmosphere is not an instantaneous process. Using a large mass ratio of $\sim$ 30 : 1 for the primary and secondary, \citet{Roediger2015c} showed that in a gradually strengthening ICM head wind, the secondary can retain a large part of its downstream atmosphere as a 'remnant tail' of unstripped gas (see Figure 3 in \citealt{Roediger2015c} for a schematic). The retained remnant tail can be larger for smaller mass ratios because the flow relaxation time and primary cluster crossing time become more equal. Thus, the secondary can retain a significant fraction of its atmosphere as it moves through the center of the primary cluster (see our images in Figure \ref{fig:slingshot_tails}, rows 1 and 2, and images of simulated mergers in, e.g., \citealt{Poole2006}; \citealt{ZuHone2011a}). \par

Remnant tails that survived pericenter passage evolve into slingshot gas tails as the secondary moves outward from the primary's center and nears the apocenter. The idea of a slingshot gas effect has been described in previous works in the context of cold fronts (\citealt{Hallman2004}, \citealt{Markevitch2007}) and gas sloshing in cluster cores (\citealt{Ascasibar2006}). \citet{Poole2006} also provided an insight into the slingshot effect and described these tails as plumes at apocenter passage, and the subsequent infall of the plume into the primary as infalling filaments. These works describe the dynamics of a slingshot tail, but focus on the formation of cold fronts and plumes rather than the characteristics of the gaseous tail. In this paper, we describe two different forms of slingshot tails, highlighting the need for caution in drawing conclusions regarding both the subhalo's direction of motion based on the tail direction, and the flow patterns surrounding slingshot tails. To this end, we analyze slingshot tails in binary cluster merger simulations, focusing on distinguishing slingshot tails from classic ram pressure stripped tails. \par

In Section \ref{sec:sims}, we outline the setup of the idealized binary cluster merger simulations we analyze in this work. In Section \ref{sec:differences} we describe the differences between a ram pressure stripped tail and a slingshot tail. Using the simulations we describe the genesis of slingshot tails and the two different forms in Section \ref{sec:slingtails}. In Section \ref{sec:flow} we describe the evolution of the flow patterns surrounding the subhalo during its journey from pericenter to apocenter, detailing how this interplays with the formation of a slingshot tail. In Section \ref{sec:obsevational} we discuss how to distinguish between ram pressure stripped tails and slingshot tails, highlighting the key observable signatures of slingshot tails. Finally, applying these insights, we identify a few known X-ray tails as slingshot tails and mark some as possible slingshot tails in Section \ref{sec:stails}. In what follows, for clarity, we term the more massive merger partner (e.g. a cluster), the primary and the less massive merger partner (e.g. subcluster or early-type galaxy), the secondary.

\section{Simulations}\label{sec:sims}
For our analysis of slingshot tails, we visually inspected the suites of idealized binary cluster merger simulations by \citet{Poole2006}, \citet{ZuHone2011a} and \citet{Sheardown2018} as well as setting up some of our own simulations for this paper based on the method detailed in \citet{Sheardown2018}. In short, all of these simulations model idealized binary cluster mergers, i.e., they set up two clusters, each in its own hydrostatic equilibrium, assign initial relative velocities to both clusters and let them collide and merge due to their mutual gravity. All simulations use the N-body method to describe the behaviour of the clusters' dark matter. This ensures dynamical friction is modelled correctly and the clusters eventually merge. It also ensures correct treatment of tidal forces. The cluster atmospheres, i.e., the ICM, is treated hydrodynamically, either by smooth particle hydrodynamics (SPH) as used in \citet{Poole2006} or by a grid method as used in \citet{ZuHone2011a, Sheardown2018}. All simulations vary the mass ratio and orbital characteristics of the merging clusters. For readers interested in more technical details we summarise those below. \par

\citet{Poole2006} present an analysis of a suite of idealized binary mergers using smoothed particle hydrodynamics (SPH) run with GASOLINE \citep{Wadsley2003}. Their simulations include the effects of radiative cooling, star formation and feedback from supernovae but neglect feedback from active galactic nuclei (AGN). The simulated clusters are idealized X-ray clusters initialised to resemble relaxed cool core clusters. The gas and dark matter properties of the clusters follow the prescription by \citet{Babul2002} and \citet{McCarthy2004}. They analyse three different cluster merger setups with mass ratios of 1:1, 1:3 and 1:10. Within each of these three setups, they run a further three sub setups which vary the initial kinematics of the secondary subhalo in concordance with the lower half of the \citet{Vitvitska2002} distribution. \citet{Vitvitska2002} showed that the average infall velocity for mergers at the virial radius is distributed normally with an average infall velocity of v$_{in}$=1.1v$_{c}$, where V$_{c}$ is the circular velocity of the secondary at the virial radius of the primary cluster. Specifically, for their three sub setups \citet{Poole2006} used values of $\upsilon_{t}$/V$_{c}$ = 0, 0.15 and 0.4, where $\upsilon_{t}$ and V$_{c}$ are the transverse and circular velocity of the secondary respectively. For the primary cluster in their simulations, the mass is set to 10$^{15}$M$_{\odot}$. \par

The simulations by \citet{ZuHone2011a} present an idealised suite of high resolution adiabatic binary cluster mergers run using FLASH, a grid based, modular hydrodynamics and N-body astrophysical code (\citealt{Fryxell2000}). The main difference between grid based and SPH codes as used by \citet{Poole2006} is there ability to resolve and handle fluid instabilities and mixing processes. While grid codes are able to do this, basic SPH methods provide poor results \citep{Agertz2007}. Furthermore, the two methods also differ in their ability to model turbulence, see \citet{Agertz2007, Dolag2005}. The mixing that will occur in the ICM due to mergers is significantly influenced by turbulence and the presence of magnetic fields. In this regard, \citet{ZuHone2011a} choose the simplest model for the ICM - an unmagnetized and inviscid gas. The N-body component of the code uses particles which simulate the behaviour of dark matter, i.e. collisionless, self gravitating particles. Including this along with the gravity associated to the gas and the gravity between both elements provides an accurate representation of tidal forces and dynamical friction during the mergers. This importantly influences the orbit of the merging subhaloes and thus the merger timescales. With FLASH, \citet{ZuHone2011a} employs the use of adaptive mesh refinement (AMR). AMR allows the user to prioritise areas of particular interest for high resolution whilst not having to use the same resolution for the whole grid. In these simulations, the authors were interested in capturing ICM shocks and cold fronts along with the inner cores of the clusters, thus high resolution is placed in these regions. Their choice of cluster initial conditions is based on cosmological simulations and observations, with the clusters initialised to be consistent with observed relaxed clusters and cluster scaling relations. More specifically, choosing clusters that lie along the M$_{500}$-T$_{X}$ relation of \citet{Vikhlinin2009}. In a similar fashion to \citet{Poole2006}, the author presents a set of three different cluster merger setups with mass ratios of 1:1, 1:3 and 1:10. Again like \cite{Poole2006}, the three merger setups each have three sub setups which are initialised with different impact parameters, but this time such that the relative tangential velocities are consistent with the \citet{Vitvitska2002} distribution. The mass of the primary cluster in this suite of simulations is set to 6 $\times$ 10$^{14}$ M$_{\odot}$. \par

%Regarding the simulation section, the general outline should be enough. Idealised clusters, binary merger, parameters varied. Maybe mass range of clusters. For specific setup you can refer to the papers

\citet{Sheardown2018} work presents three tailored simulations of the infall of the elliptical galaxy NGC 1404 into the Fornax Cluster. As with \citet{ZuHone2011a}, their simulations were run using FLASH, using a similar simulation design. Their simulations did not include the effects of radiative cooling or heating by AGN. The inclusion of both these features would only affect the properties of the gas in the very central regions of the cluster and the galaxy, and as the authors report, their results did not rely on the central gas cores. Each simulation differed by the initial kinematics of NGC 1404, i.e., the secondary. One simulation starts with NGC 1404 having an almost zero infall velocity, with just a small tangential component to ensure that the merger is not a head on collision (as this was ruled out by observation). For the other two simulations the initial velocity is set to v$_{in}$=1.1v$_{c}$, the average infall velocity at the virial radius in accordance with \citet{Vitvitska2002}. They then differ by the initial tangential velocity component which is set in agreement with \citet{Vitvitska2002}. The mass of the Fornax Cluster is set at 6 $\times$ 10$^{13}$ M$_{\odot}$ and for NGC 1404, 0.45 $\times$ 10$^{13}$ M$_{\odot}$, making it $\sim$ 1:10 merger. \par

The 1:3 merger shown in Figures \ref{fig:slingshot_tails}, \ref{fig:slingshot_proj} and \ref{fig:flow2} we ran for the purposes of this paper. The primary and secondary are modelled such that they follow the setup procedure in \citet{ZuHone2011a} but use a Hernquist profile for the total mass distribution. The simulation design follows \citet{Sheardown2018}. The mass of the primary is set to 6 $\times$ 10$^{14}$M$_{\odot}$ and the initial velocity of the secondary follows the \citet{Vitvitska2002} condition, v$_{in}$=1.1v$_{c}$ using a tangential velocity of v$_{\perp}$=0.71v$_{c}$. The 1:1 merger shown in Figure \ref{fig:ngc7618} we ran to provide a visual match to the observed image of NGC 7618 and UGC 12491. This is the same simulation as described in \citet{Sheardown2018}, using the setup for the cluster which has a mass of 6 $\times$ 10$^{13}$ M$_{\odot}$ and using an initial tangential velocity component of v$_{\perp}$=0.71v$_{c}$.

\section{Ram Pressure Stripped Tail vs Slingshot Tail} \label{sec:differences}

To begin with, it is important that we affirm the difference between a ram pressure stripped tail and a slingshot tail. A ram pressure tail is formed due to the motion of the secondary against the ICM of the primary, where the ram pressure is equal to P$_\text{ram}$ $\approx$ $\rho_\text{ICM}$v$_\text{sec}^{2}$, where v$_\text{sec}$ is the velocity of the secondary with respect to the ICM of the primary. During the infall phase, the increasing ram pressure progressively strips the gaseous atmosphere of the secondary into a downstream tail which points directly opposite to the direction of motion, producing an orderly head-tail structure, as demonstrated in e.g. \citet{Acreman2003} and \citet{Roediger2015c}. The part of the gas tail closest to the secondary is a remnant tail, i.e., the still unstripped, bound downstream atmosphere of the secondary that is shielded from the upstream ICM wind, as shown in Figure \ref{fig:slingshot_tails}, row 1. In the frame of the secondary, the flow of the primary's ICM around the secondary closely follows the classic flow around a blunt body, including an upstream stagnation point, strong shear flow along the sides of the secondary and a downstream deadwater region as the start of a long wake. \par 

\begin{figure}
\centering
\includegraphics[scale=0.4]{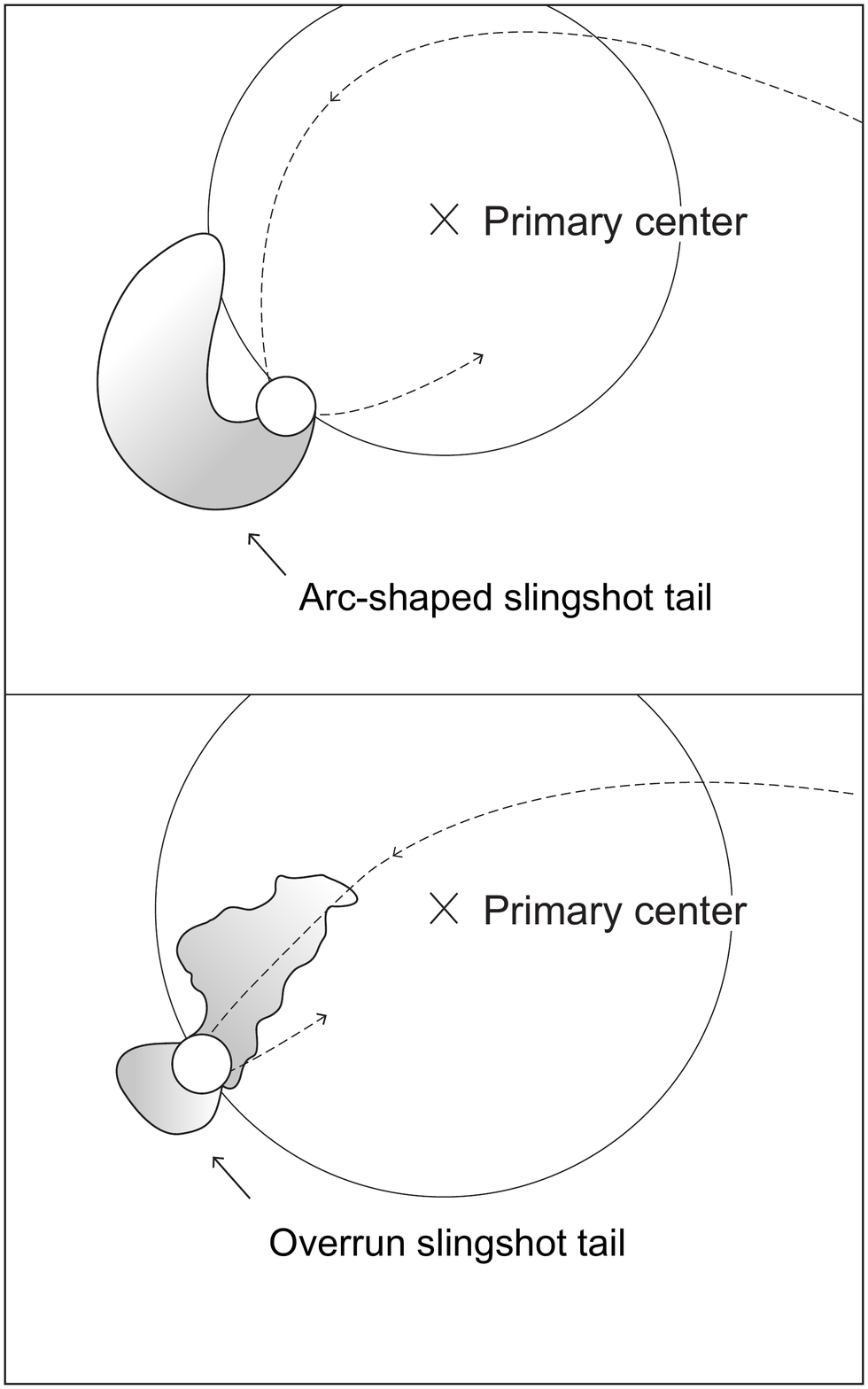}
\caption{A cartoon image showing a clear distinction between the two slingshot tail forms. The primary center is marked with a cross and the small white circle represents the secondary at apocenter. The dashed line represents the approximate orbit of the secondary. In the arc-shape slingshot form, a prominent arc-shaped tail is produced when the secondary reaches apocenter. For the overrun slingshot form, the tail slingshots directly over the secondary producing an irregular shaped atmosphere followed by a conical shape tail behind it.}
\label{fig:slingshot_cartoon}
\end{figure} 

The dynamics change when the secondary has passed the pericenter and moves toward the next apocenter of its orbit. Now the ram pressure on the secondary's atmosphere rapidly decreases due to its decreasing velocity and the decrease in ICM density. As the secondary slows and eventually turns around, the still bound gas from the ram pressure tail falls back toward the secondary's center due to the secondaries gravity and overshoots it in a slingshot effect, resulting in a slingshot tail that can point sideways or even opposite to the direction of motion of the secondary, contrasting with the orderly head-tail structure of a ram pressure stripped tail. Additionally, during the formation of the slingshot tail, the ICM flow around the secondary does not follow the flow around a blunt body any more but becomes highly irregular, as detailed further in Section \ref{sec:flow}. Along with this gas dynamics effect, tidal decompression of the secondary after pericenter passage plays a role too in the shaping of the tail, similar to the long tidal tails created in pure N-body mergers. Adiabatic expansion makes the tail cooler too as it is sling-shotted into the lower pressure ICM environment. In short, around the apocenter of the orbit, the secondary is not being ram pressure stripped and the tail has been shaped by tidal forces more than just the ram pressure. Therefore a tail observed in the slingshot state should not be identified as a gas stripping tail as this scenario does not accurately describe the physics of the situation. The application of the classic ram pressure stripping scenario to a slingshot tail will lead to incorrect conclusions in regard to the subhalo velocity, expected locations of shear flows, instabilities and mixing (detailed in Section \ref{sec:flow}). For example, as the slingshot tail can point sideways or ahead of the subhalo, it does not trace the recent orbit path like an orderly ram pressure stripped tail would, and is therefore misleading when drawing naive conclusions regarding the direction of the subhalo based on the tail direction. \par

As mentioned, \citet{Hallman2004} described a ram pressure slingshot mechanism to explain the cold front which appears ahead of the northern subcluster in the merging cluster A168. This idea has further been used to describe merger features in Abell 2744 also (\citealt{Owers2011a, Merten2011}). The formation of these cold fronts found ahead of the subcluster were predicted in hydrodynamical simulations by \citet{Mathis2005} and by \citet{Ascasibar2006} in the context of gas sloshing. These slingshot cold fronts are the contact discontinuity between the slingshot tail and the primary's ICM. \par
 
\begin{figure*}
\centering
\includegraphics[scale=0.85]{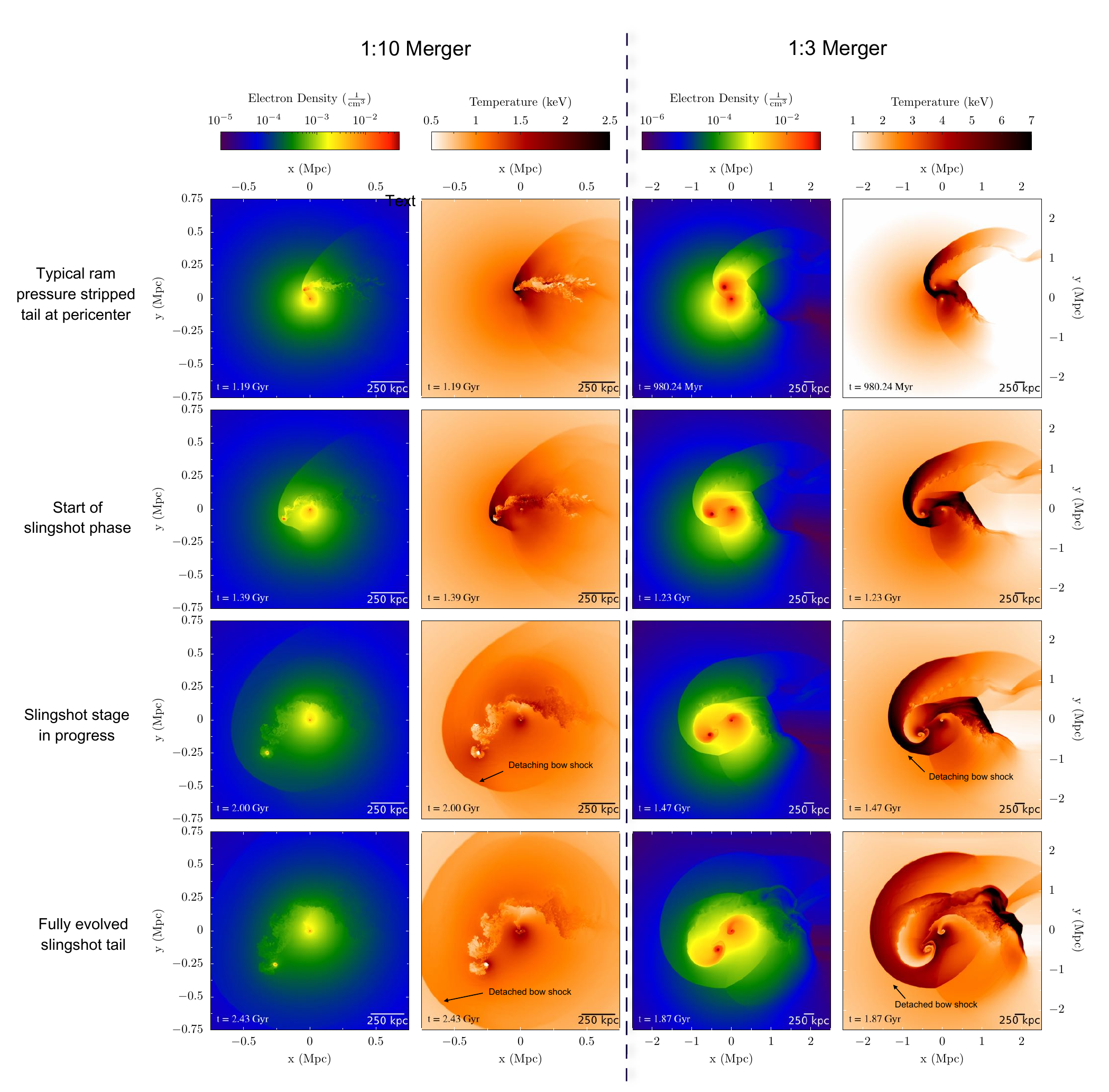}
\caption{Evolution of two different cluster mergers. The first and second column show electron density and temperature slices for a $\sim$ 1:10 merger with a small impact parameter from the V1 simulation in \citet{Sheardown2018}. The third and fourth column likewise show electron density and temperature slices but for a 1:3 merger we ran for this paper (following the simulation design of \citet{Sheardown2018}) using two idealized clusters with a pericenter distance of 330 kpc. The first row shows the secondary at pericenter with a typical ram pressure stripped tail. Note that in the 1:3 merger, the secondary still contains a large amount of unstripped gas. The second row shows the start of the slingshot tail being produced as the secondary slows toward apocenter. In the third row, for the 1:10 merger, the first phase of the overrun slingshot form is established, with the secondary harboring an irregular shaped atmosphere as the remnant tail overruns directly the remnant atmosphere. For the 1:3 merger, the arc-shaped tail becomes a prominent feature. In the fourth row, for the 1:10 merger, the second phase is reached as the remnant tail continues to overrun the remnant atmosphere and fans out along the direction of apocenter away from the secondary. For the 1:3 merger, the arc-shaped tail reaches its full prominence as the secondary turns around and begins to infall again. In the fourth row, we also mark the bow shock that detaches from the slowing down secondary. The detached bow shock will continue moving away from the primary's center.}
\label{fig:slingshot_tails} 
\end{figure*}

%% The reader needs to have the characteristics of the simulations presented in detail in a dedicated section, with assumptions, limitations, strengths and weakness points clearly presented

\section{Slingshot Tails}\label{sec:slingtails}
We find that as long as the secondary can retain some remnant tail through pericenter passage, it develops a slingshot tail. We further find that slingshot tails can be split into two main distinct forms, each giving characteristically different morphologies dependent on the impact parameter and mass ratio of the merger. We term these forms \emph{arc-shaped slingshot tails} and \emph{overrun slingshot tails}. For both forms, we find that the slingshot tail stage typically lasts between 0.5 - 1.0 Gyr, thus slingshot tails may be fairly common since the secondary spends much more time around apocenter than during pericenter passage, where they are moving faster. Figure \ref{fig:slingshot_cartoon} presents a cartoon image of both cases to provide a clear visual distinction between the two. We note that there are some cases which do not fall cleanly into one of these forms and are somewhere in between, in this paper we only focus on the two extreme cases of the \emph{overrun} or \emph{arc-shaped} form. We find that lower mass ratio mergers tend to result more in the overrun form, however as seen in Figure 11 in \citet{ZuHone2011a}, a 1:10 mass ratio with large impact parameter results in an arc-shaped slingshot tail, so this is not always the case. In addition to the impact parameter and mass ratio, bulk motions of the ICM in the primary cluster, triggered by the merger also, play a significant role in the evolution of the secondary's slingshot tail. Deriving the exact conditions for one or the other slingshot forms requires a separate, more systematic study. In the following we describe the generation of the two main slingshot forms and discuss their underlying physics. We remind the reader that we are now concerned with the merger phase where the secondary moves from pericenter toward apocenter and starts its next infall.

\subsection{Arc-Shaped Tails}\label{sec:mode1}
\begin{figure*}
\centering
\includegraphics[scale=0.72]{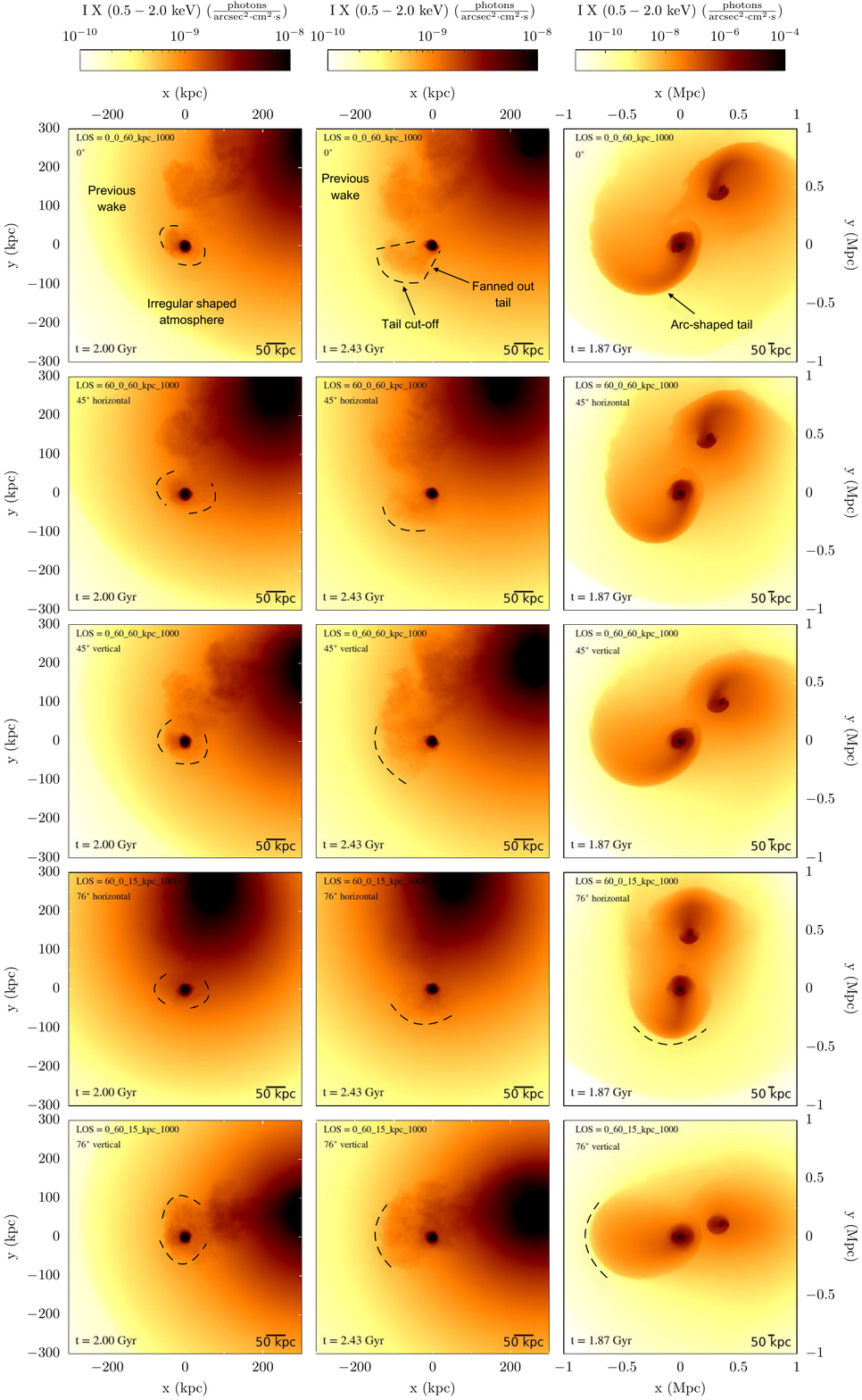}
\caption{X-ray photon intensity field projections calculated in the 0.5-2.0 keV energy band for the different slingshot tail forms as shown in Figure \ref{fig:slingshot_tails}. The first column shows the overrun slingshot tail in the first phase, where the secondary harbors an irregular shaped atmosphere. The second column shows the overrun slingshot tail in the second phase, where the secondary possesses a fanned out tail. The third column shows the arc-shaped slingshot tail. The top row is a projection perpendicular to the orbital plane. The following images are a selection of LOS rotated vertically or horizontally to the orbital plane by 45$^{\circ}$ and 76$^{\circ}$. Each image is annotated with its corresponding rotation and angle. Crucially, we see that regardless of projection angle, the features of both slingshot forms remain intact. For the arc-shaped form, the tail remains prominent but for certain angles it may not appear as arc-like. To distinguish this case from a ram pressure stripped tail would be the homogeneous brightness of the tail along with its distinct downstream edge.}
\label{fig:slingshot_proj}
\end{figure*}

When the impact parameter of the merger is large, the remnant tail of the secondary, that was once pointing downstream (toward the direction of pericenter), is carried out sideways, by angular momentum conservation, to the side of the secondary furthest from the primary cluster center as it approaches apocenter. This results in a prominent arc-shaped tail that can point sideways to the secondary as shown in Figure \ref{fig:slingshot_tails}, columns 3 and 4, and in Figures, 5, 8 and 11, snapshot 2.0 Gyr in \citet{ZuHone2011a}. The archetypal arc-shaped slingshot tails tend to consist largely of still unmixed, cool gas that always belonged to the secondary. Due to the absence of internal shear, these tails also tend not to be turbulent. Shear and the resulting Kelvin-Helmholtz instabilities (KHI) appear mainly along the far end or the outer wing of the arc-shaped slingshot (see also Section \ref{sec:flow}). The size of the arc-shape tail is very much dependent on the impact parameter and initial gas contents, as this generally dictates the amount of gas the secondary can carry through pericenter passage. Naturally, the larger the impact parameter, the larger the tail, as the stripping due to ram pressure will not be as strong, hence more gas can be retained. Therefore, the size of the arc-shape tail can potentially be used to infer the impact parameter for the merger. We also find that the arc-shaped slingshot tails can 'swing' all the way around from one side of the secondary as it approaches apocenter, to the other side as it moves through apocenter to the beginning of the next infall. Furthermore, when the masses of the merging systems are similar, we see that the primary develops a slingshot tail that is similar in size to the secondary's, appearing symmetric. In Figure \ref{fig:slingshot_proj}, column 3, we present a variety of X-ray projections for the arc-shaped slingshot tail form. Most features of the tail do not change depending on the viewing angle. The tail remains homogeneous in brightness and has a sharp edge away from the merger companion. These edges have been called slingshot cold fronts previously. When we see the plane of the merger almost edge-on, the arc-shape slingshot tail may not point sideways, in this scenario the homogeneous brightness and sharp edge of the tail can be used to distinguish from a ram pressure stripped tail.

\subsection{Overrun Tails}\label{sec:mode2}
In contrast to arc-shaped slingshot tails, when the impact parameter is small, the remnant tail slingshots directly along the outgoing orbit and, as the secondary reaches apocenter, it is overrun by its own slingshot tail. We term this \emph{overrun slingshot tail}. Its evolution can be split into two distinct phases. As the secondary decelerates due to the gravitational pull of the primary cluster and dynamical friction, the lower orbital angular momentum causes the gaseous tail to overshoot directly over the potential center of the secondary. This creates the first phase where the secondary appears to harbor a second gas atmosphere which encompasses the secondary's true remnant atmosphere, resulting in an overall 'fuzzy' irregular shape as shown in Figure \ref{fig:slingshot_tails}, column 1 row 3 and in Figure \ref{fig:slingshot_proj}, column 1. This secondary atmosphere is turbulent (assuming no other processes suppress turbulence) as the remnant tail continues to overrun the secondary. This feature can also be seen in the simulations of \citet{Acreman2003}, specifically in their Figure 2d. In the second phase, the actual slingshot tail appears as a conical shaped tail which progressively fans out along the direction pointing away from the primary cluster center, as shown in Figure \ref{fig:slingshot_tails}, column 1 row 4 and Figure \ref{fig:slingshot_proj}, column 2. The overrun slingshot tail is likely always turbulent as there are more locations with shear flows. In result, the overrun tail is well mixed with the ambient ICM, and appears homogeneous in both density and surface brightness where both lie in between that of the ambient ICM and the remnant core of the secondary. We also find that the fanned out tail in the second phase is cut off on the far side away from the cluster center, in a similar manner to the arc-shaped slingshot tails. This cut off point marks the maximum radius the tail slingshots to. For this form, the next infall of the secondary occurs almost along the path of its previous wake due to the lower orbital angular momentum. In Figure \ref{fig:slingshot_proj}, columns 1 and 2, we show X-ray projections for a variety of viewing angles for both phases of the overrun slingshot tail. Regardless of the viewing angle, the characteristic features of the overrun form remain clear. \par

\section{Flow Patterns of Slingshot Tails} \label{sec:flow}

\begin{figure*}
\centering
\center{\emph{Overrun slingshot tail}}\\
\begin{minipage}{0.91\textwidth}
\centering
\begin{minipage}{0.26\textwidth}
\centering
\includegraphics[trim={0cm 1.0cm 1.7cm 0cm}, clip,width=\columnwidth]{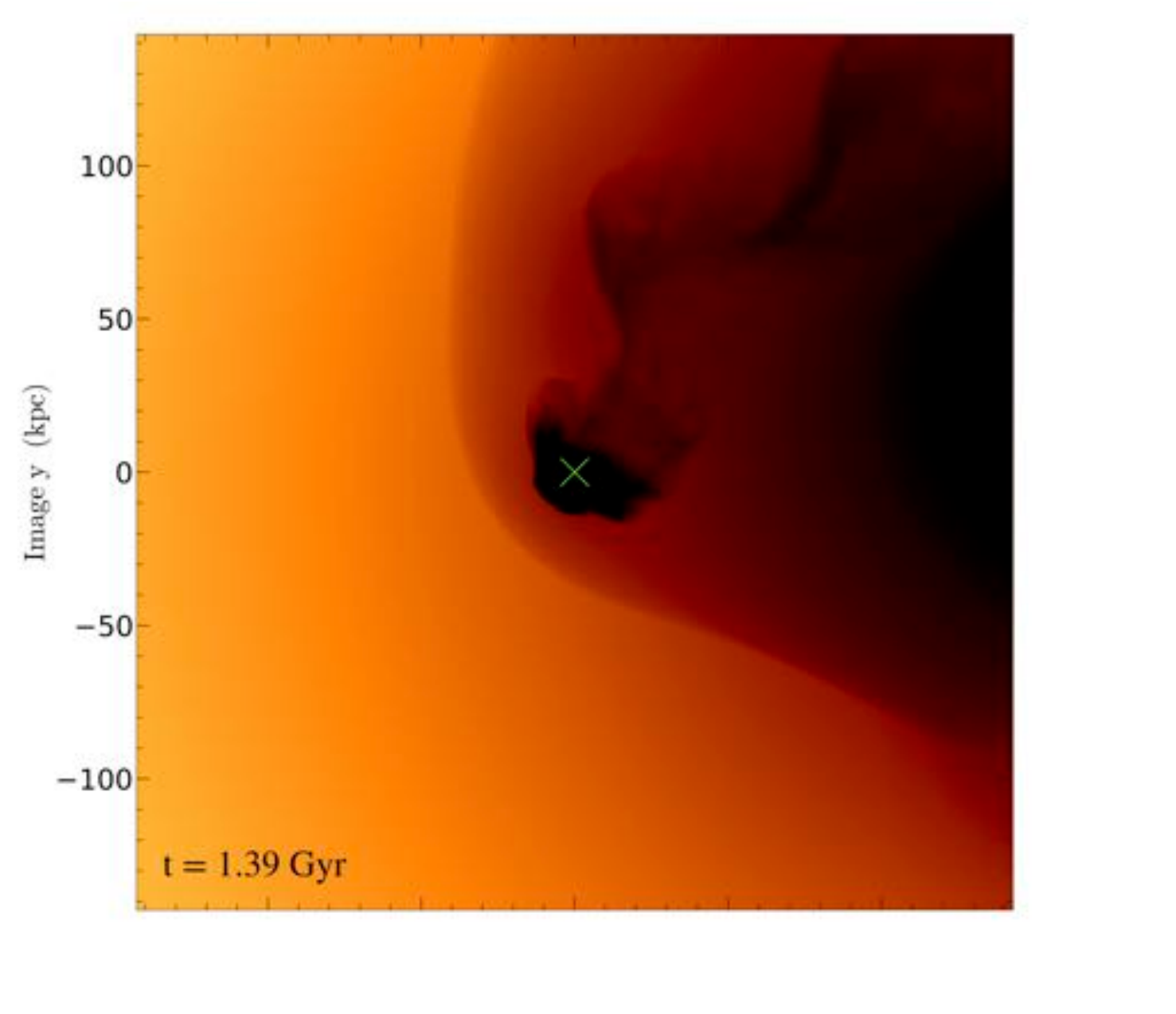}
\includegraphics[trim={0cm 1.0cm 1.7cm 0cm}, clip,width=\columnwidth]{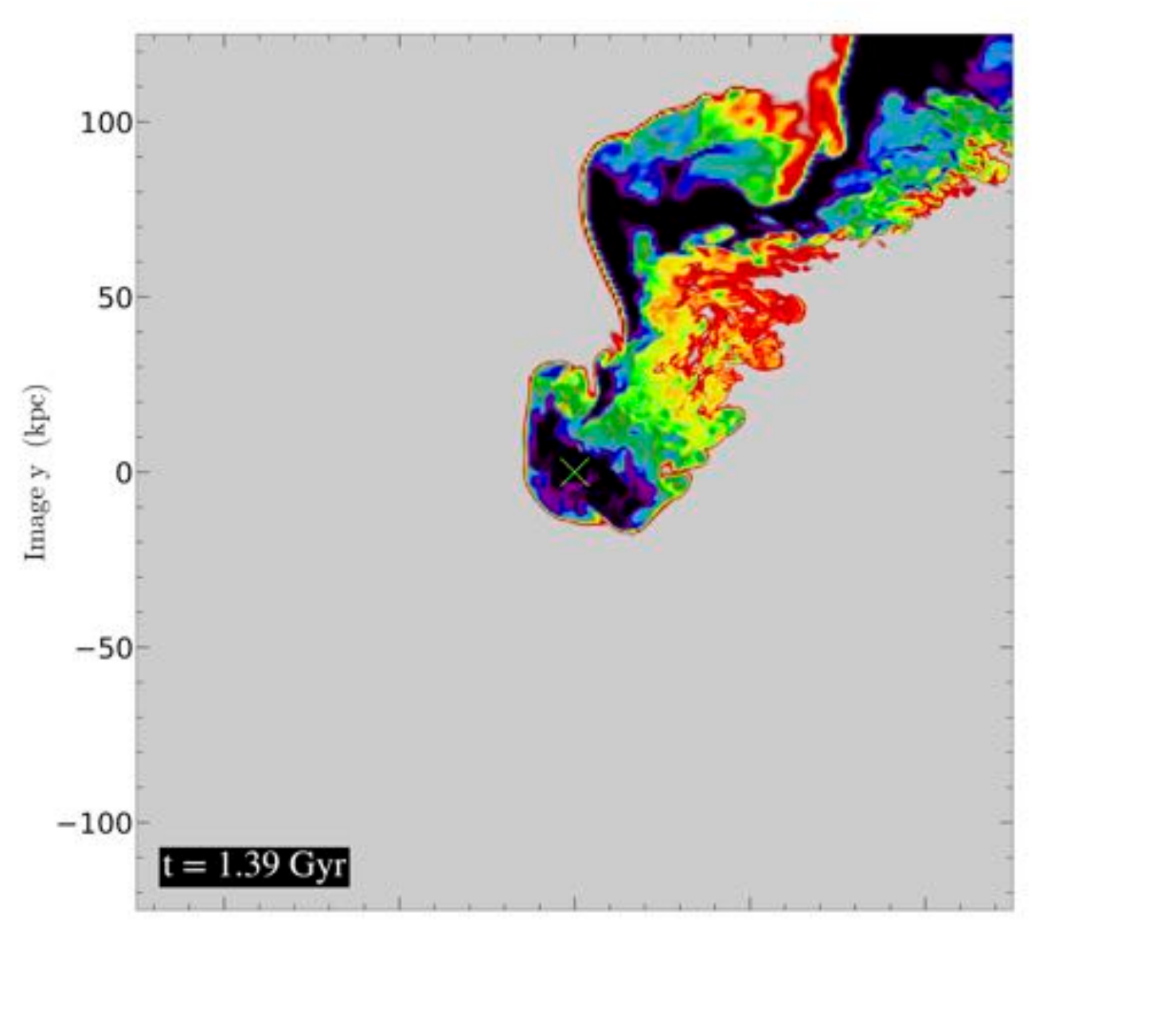}
\includegraphics[trim={0cm 1.0cm 1.7cm 0cm}, clip,width=\columnwidth]{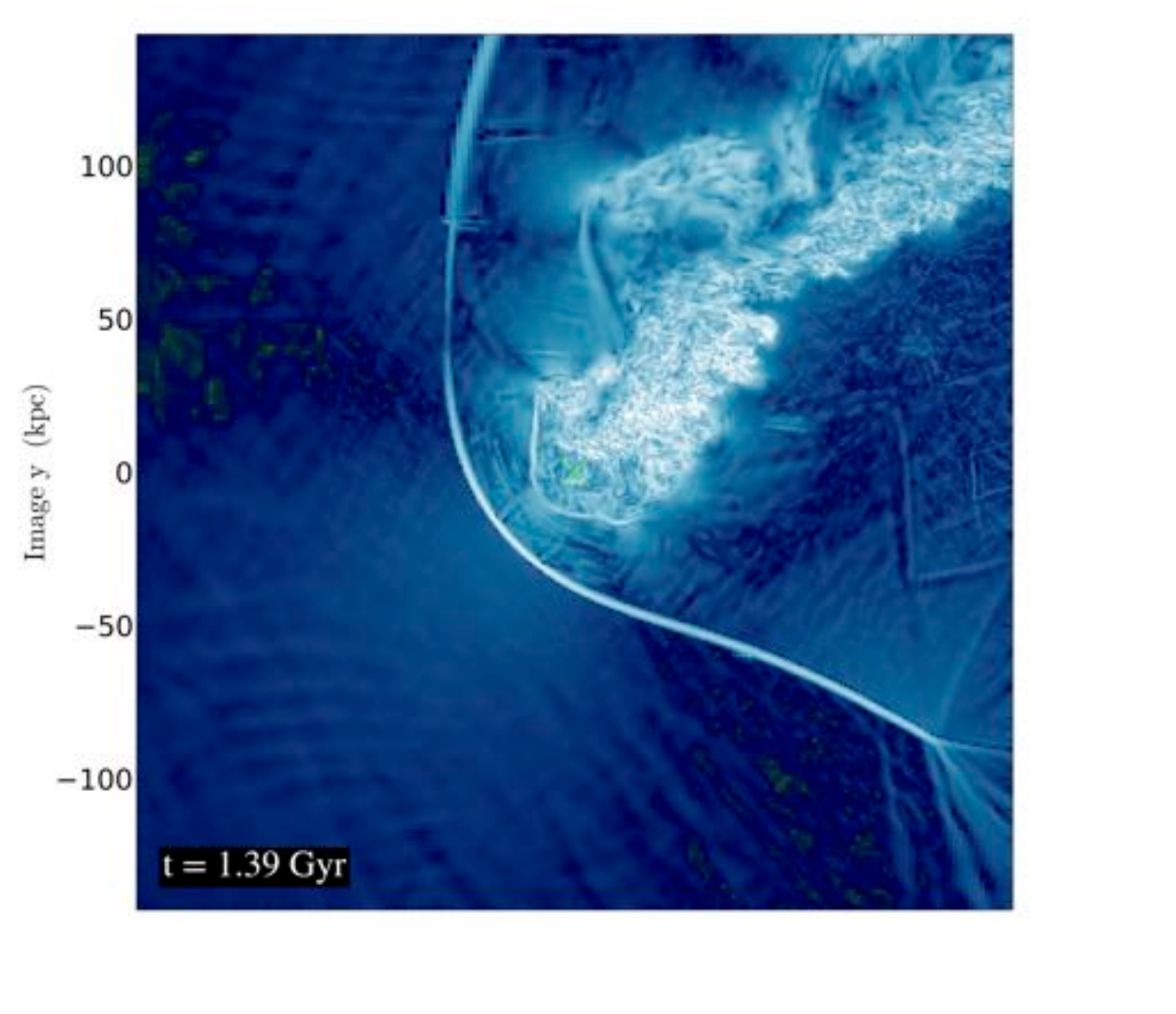}
\includegraphics[trim={0cm 0.5cm 1.7cm 0cm}, clip, width=\columnwidth]{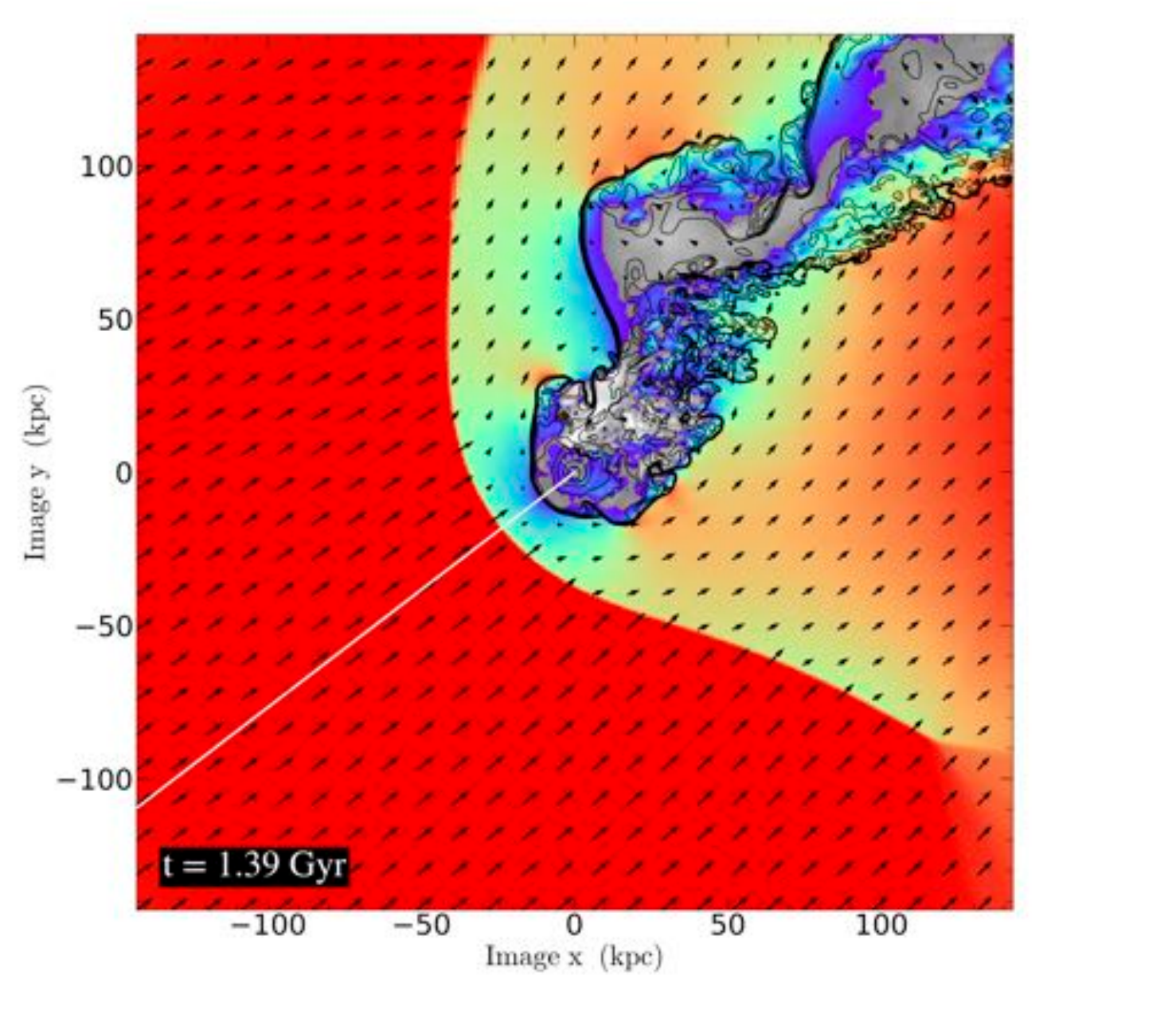}
\\(a)\\
\end{minipage}
\begin{minipage}{0.26\textwidth}
\centering
\includegraphics[trim={1.7cm 1.0cm 1.7cm 0cm}, clip, width=0.873\columnwidth]{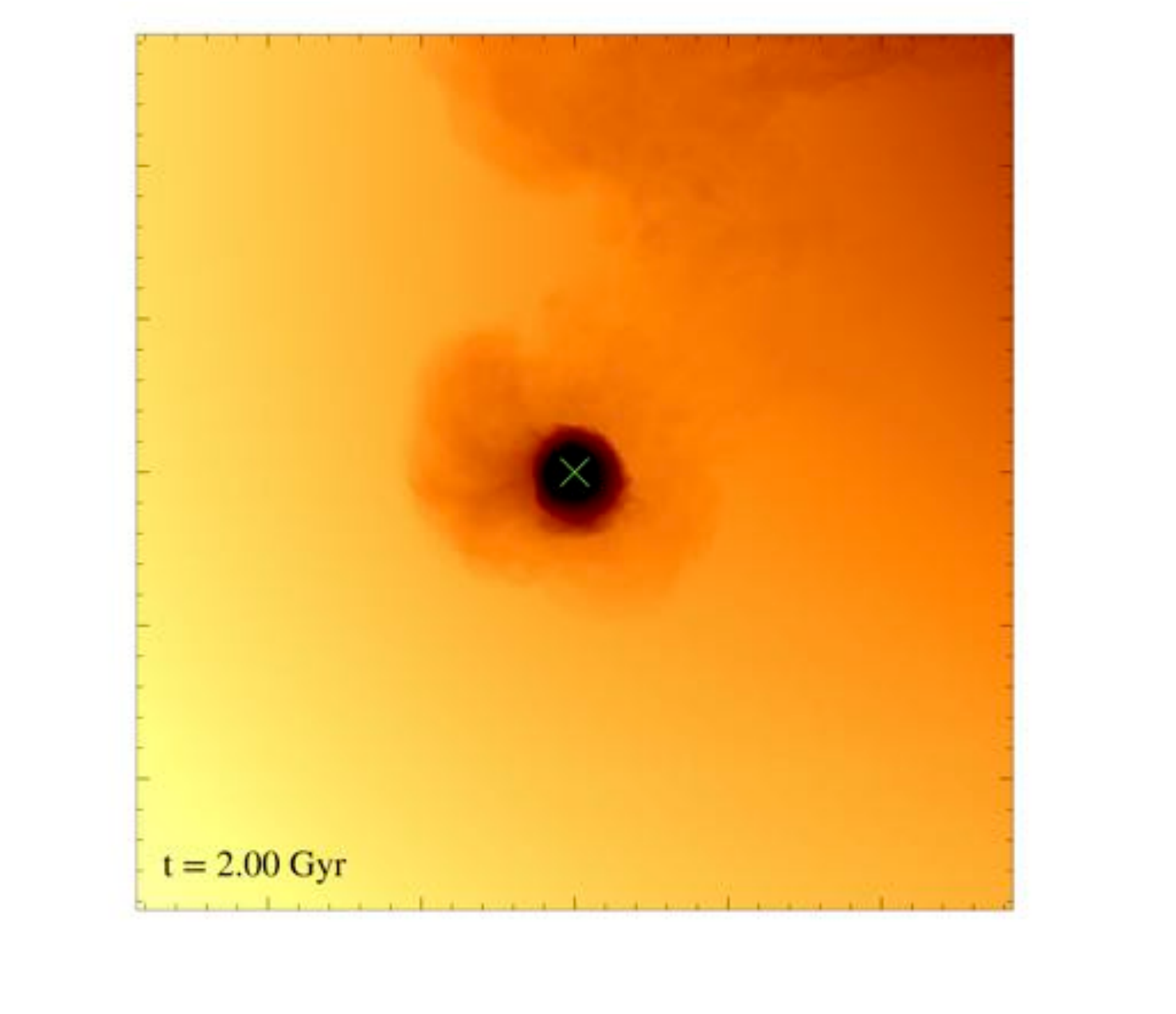}
\includegraphics[trim={1.7cm 1.0cm 1.7cm 0cm}, clip, width=0.873\columnwidth]{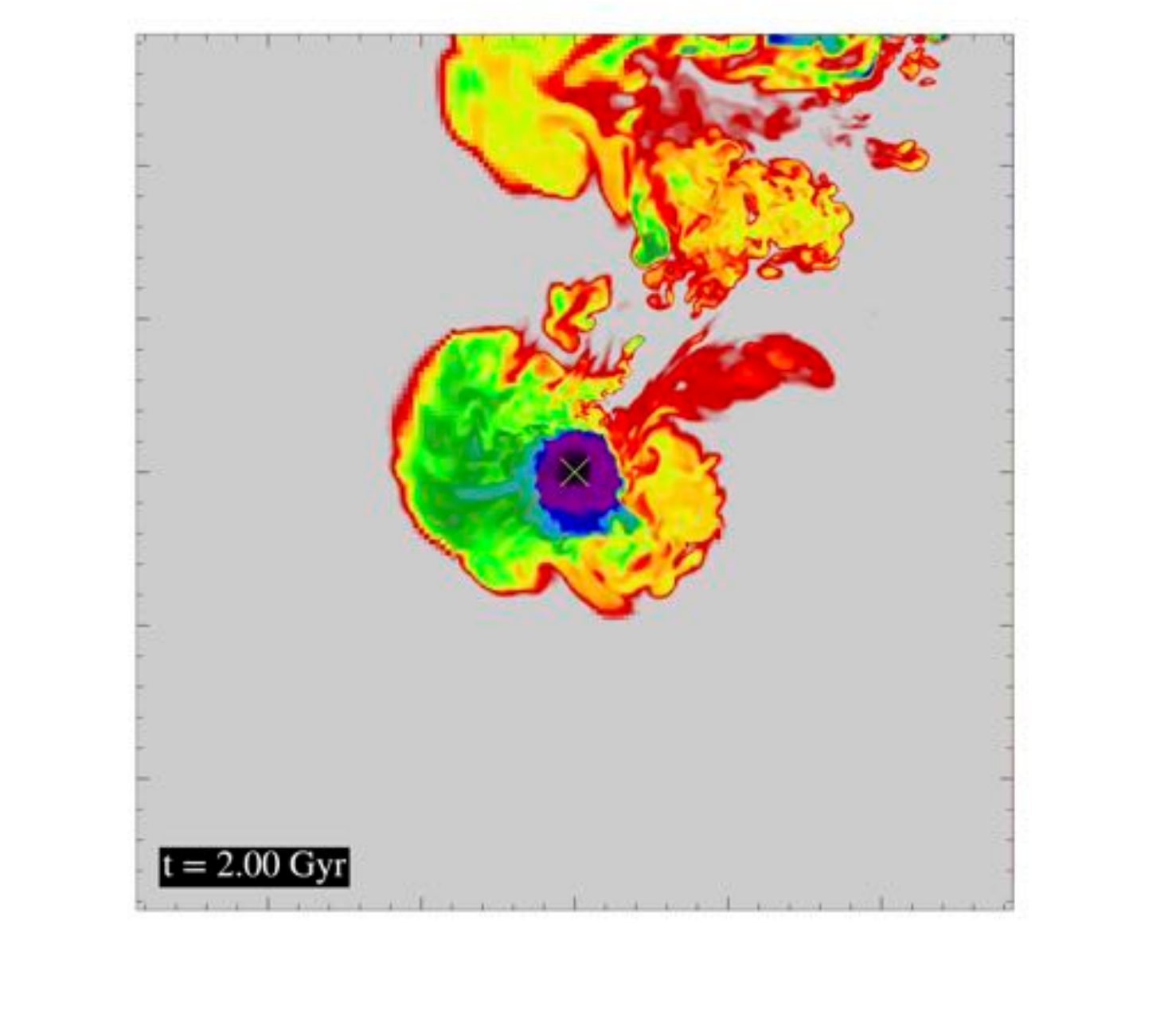}
\includegraphics[trim={1.7cm 1.0cm 1.7cm 0cm}, clip, width=0.873\columnwidth]{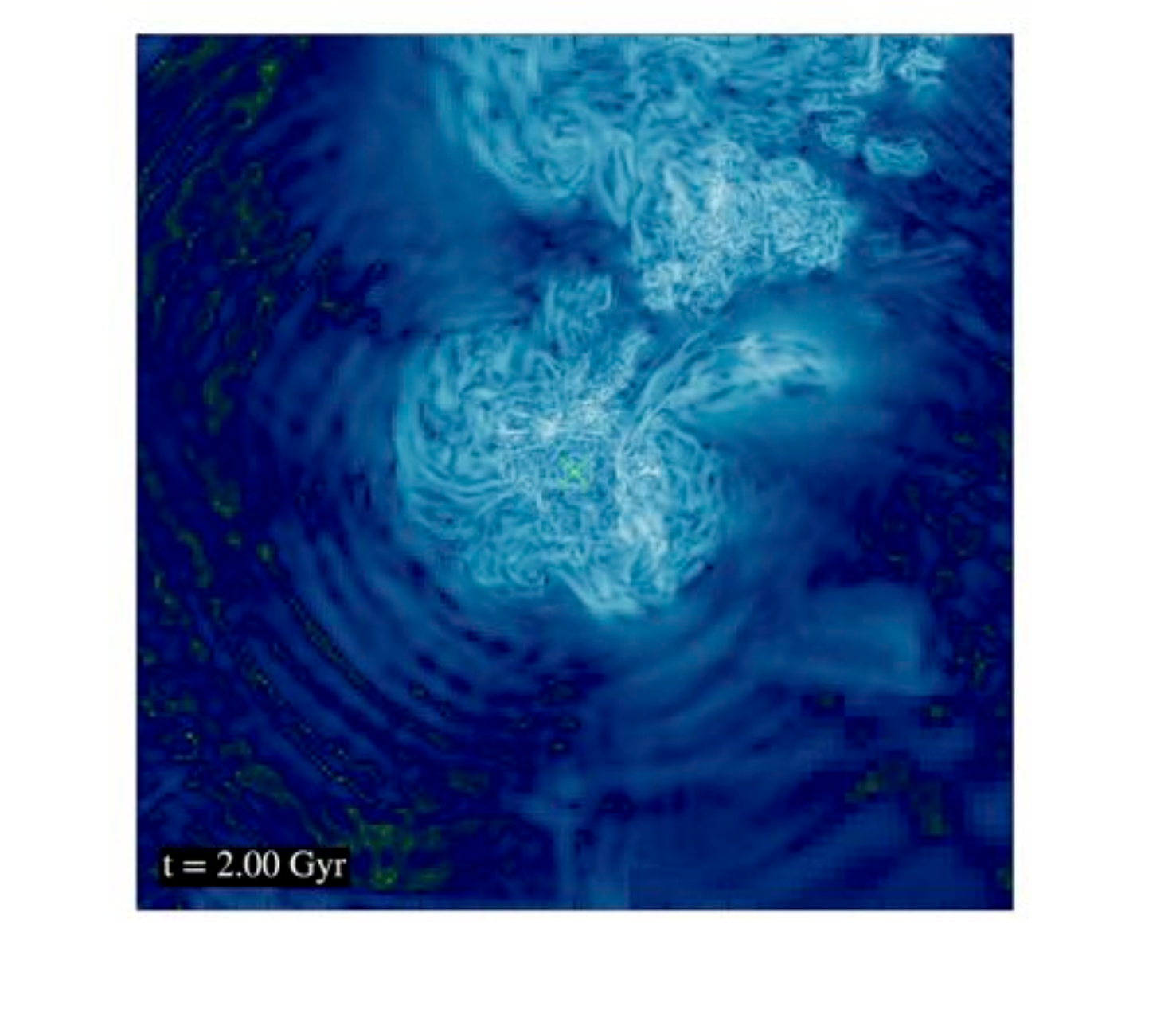}
\includegraphics[trim={1.7cm 0.5cm 1.7cm 0cm}, clip,width=0.873\columnwidth]{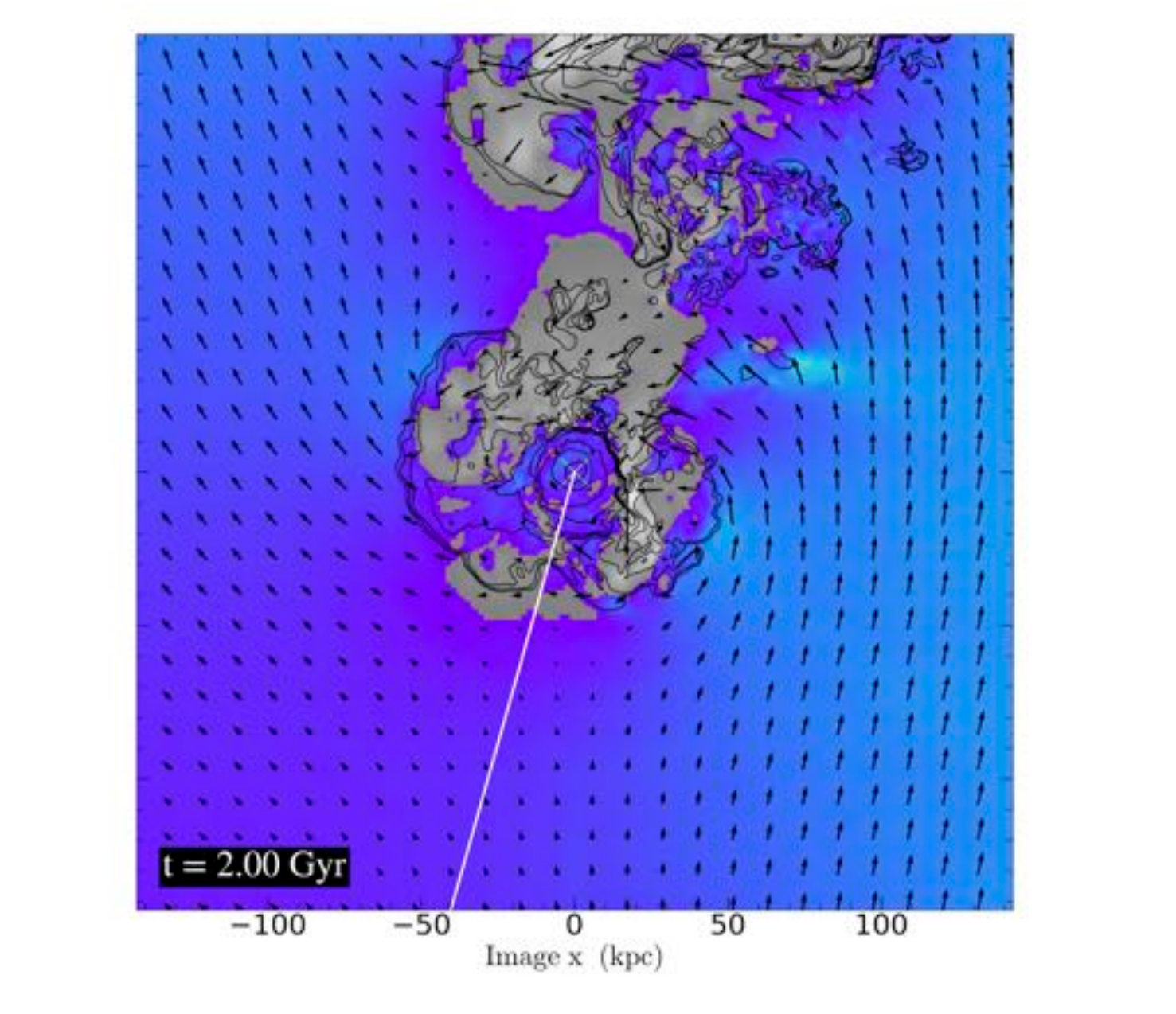}
\\(b) Phase 1\\
\end{minipage}
\begin{minipage}{0.26\textwidth}
\centering
\includegraphics[trim={1.7cm 1.0cm 0cm 0cm}, clip,width=\columnwidth]{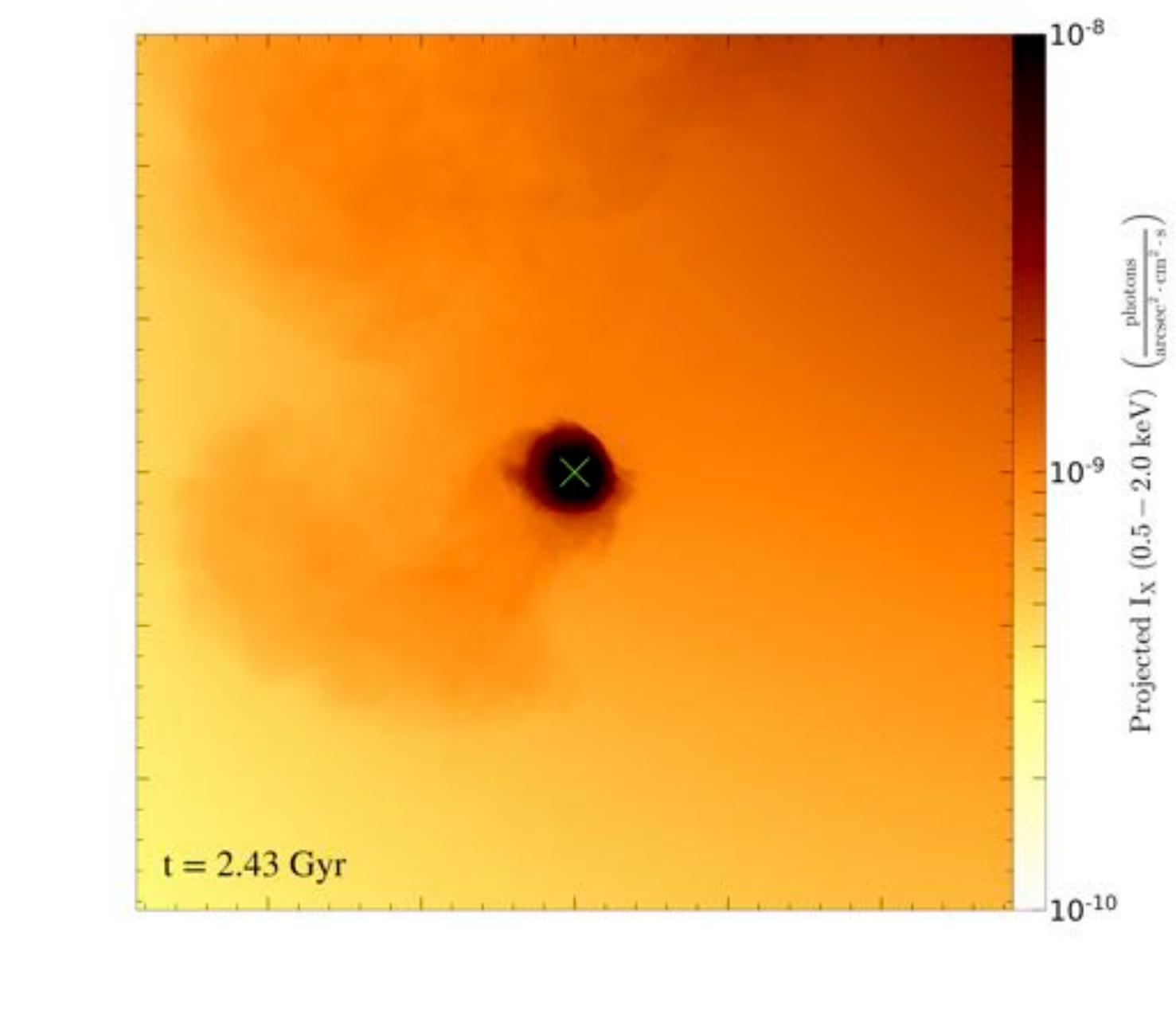}
\includegraphics[trim={1.7cm 1.0cm 0cm 0cm}, clip,width=\columnwidth]{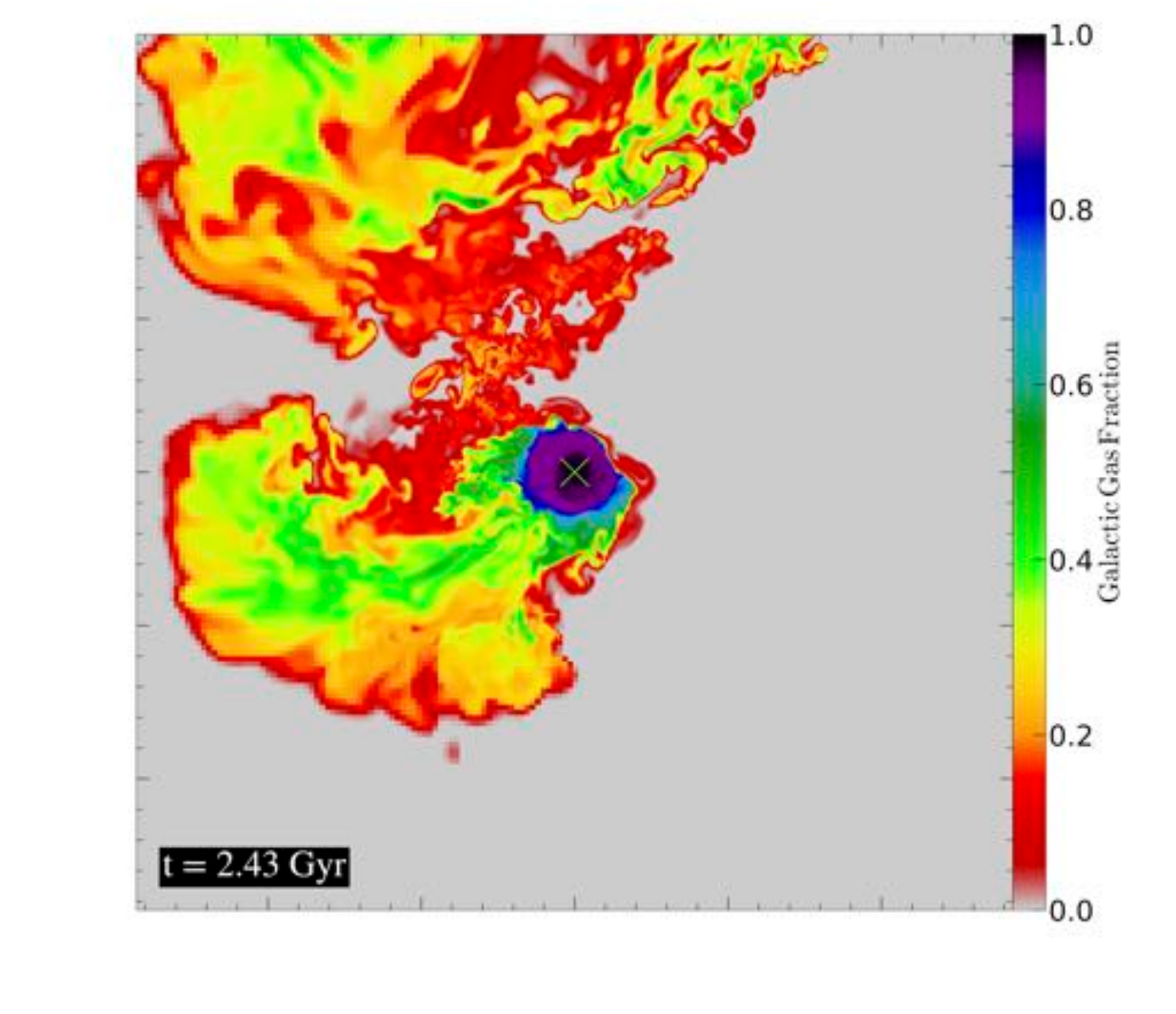}
\includegraphics[trim={1.7cm 1.0cm 0cm 0cm}, clip,width=\columnwidth]{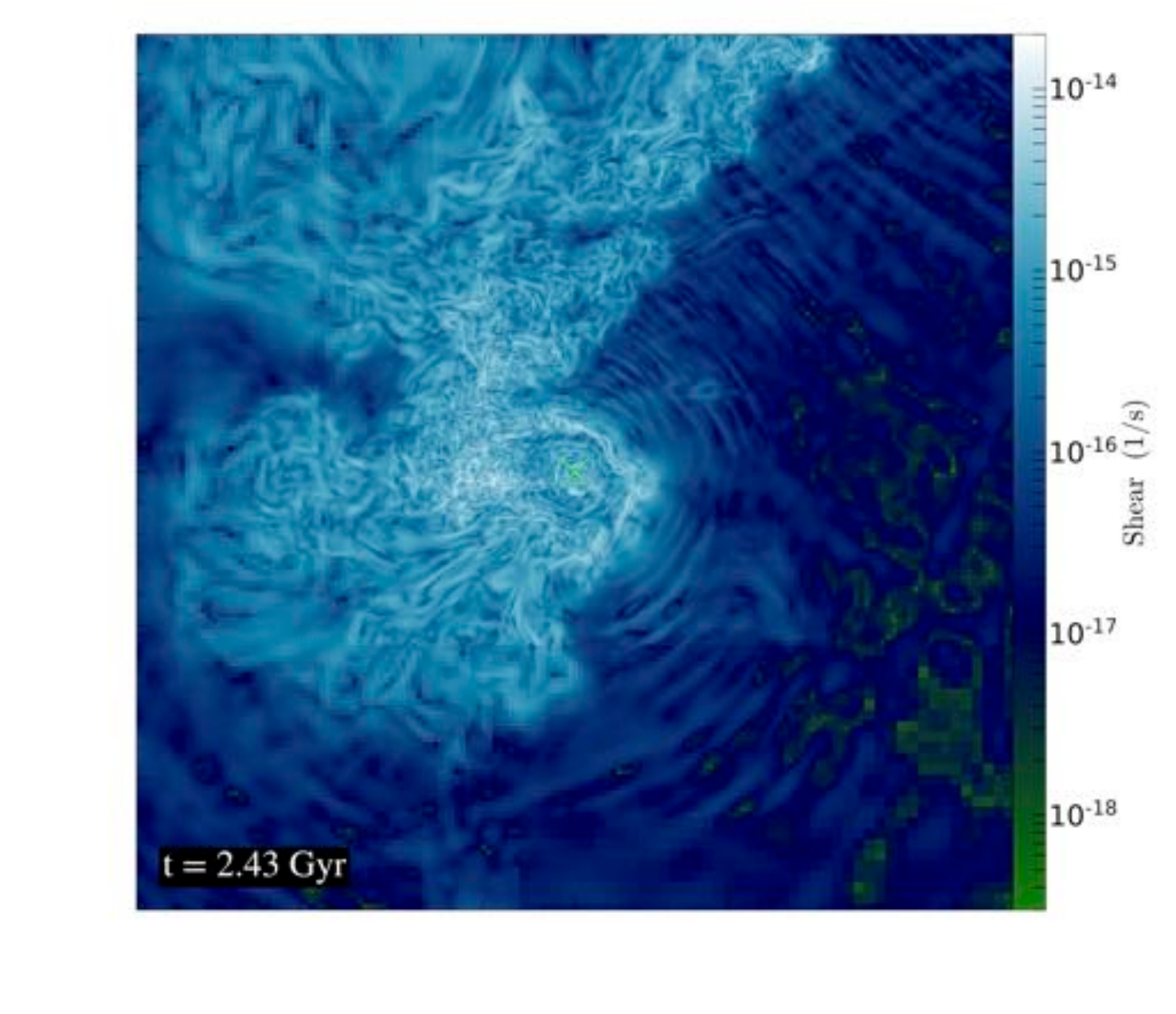}
\includegraphics[trim={1.7cm 0.5cm 0cm 0cm}, clip,width=\columnwidth]{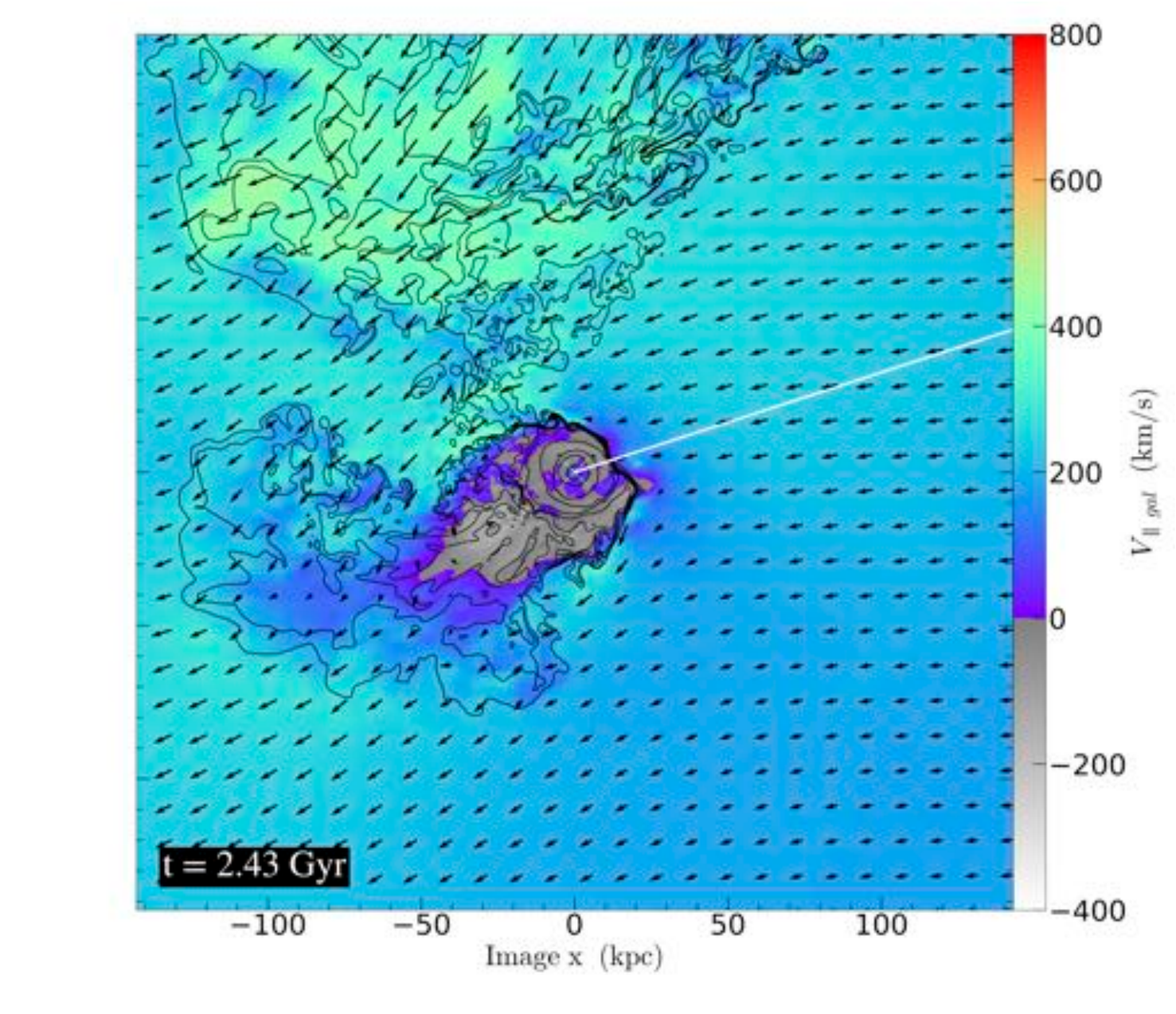}
\\(c) Phase 2\\
\end{minipage}
\\
(See \S\ref{sec:mode2} for more on Phases 1 \& 2)
\end{minipage}
\centering
\caption{The aim of this figure is to show the flow patterns in and around secondaries with \emph{overrun slingshot tails}. The images are made from the V1 simulation in \citet{Sheardown2018}, the same as the two left-hand columns of Figure~\ref{fig:slingshot_tails}. Each column shows an X-ray photon intensity projection in the orbital plane; a gas fraction slice of the secondary, showing the extent that the tail has been stripped and mixed with the ICM; a slice of the shear rate, showing the locations of strong shear flows; and finally a colormap of the flow field, overlaid with velocity vectors. For the latter, the colormap codes the velocity component $V_{\parallel \,\, gal}$ parallel to the secondary's direction of motion, in the rest frame of the secondary. The white line from the secondary center shows the direction of motion of the secondary with respect to the grid and the contours show the gas density of the secondary's atmosphere as it is stripped. The rainbow part of the colormap shows gas flow toward the secondary's downstream direction, while the gray scale part shows the flow toward the upstream direction.\\
\\
The images in Column (a) show the unstable flow beginning to develop. (b) shows the secondary near apocenter as the \emph{overrun slingshot tail} is in the first phase with an irregular shaped atmosphere. (c) shows the flow shortly before it becomes classed by this paper as a stable infall again, where now we have phase two of the \emph{overrun slingshot tail} as a conical tail develops behind the secondary. Both (a) and (c) can both be considered fringe cases in terms of the flow stability. This figure demonstrates that during the creation of a slingshot tail, the secondary undergoes a significant asymmetrical flow relative to its direction of motion - even in the case of (c), which may be considered steady based on X-ray observations.}
\label{fig:flow}
\end{figure*}

\begin{figure*}
\center{\emph{Arc shaped slingshot tail}}\\
\centering
\begin{minipage}{0.91\textwidth}
\centering
\begin{minipage}{0.26\textwidth}
\centering
\includegraphics[trim={0cm 1.0cm 1.7cm 0cm}, clip,width=\columnwidth]{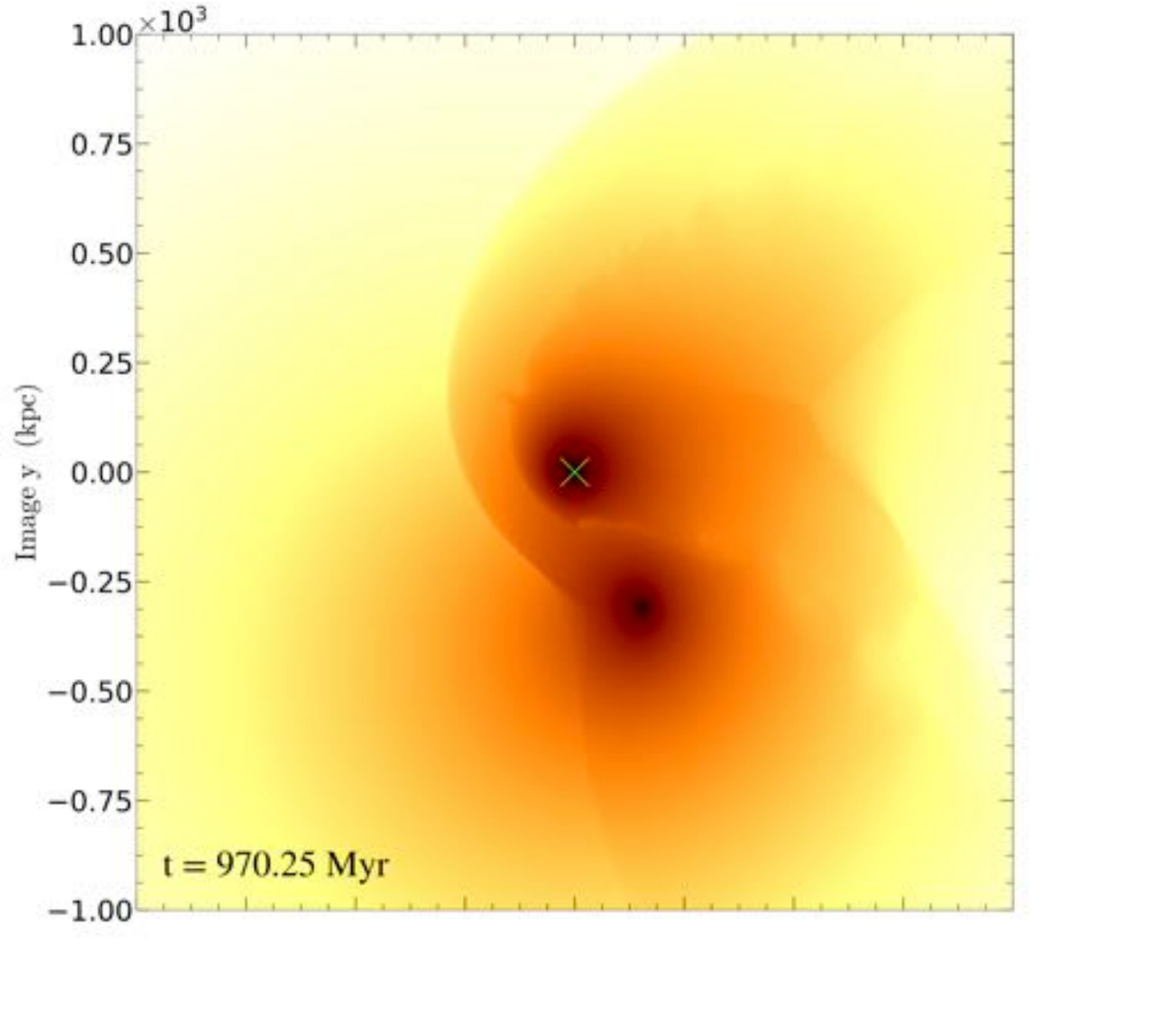}
\includegraphics[trim={0cm 1.0cm 1.7cm 0cm}, clip,width=\columnwidth]{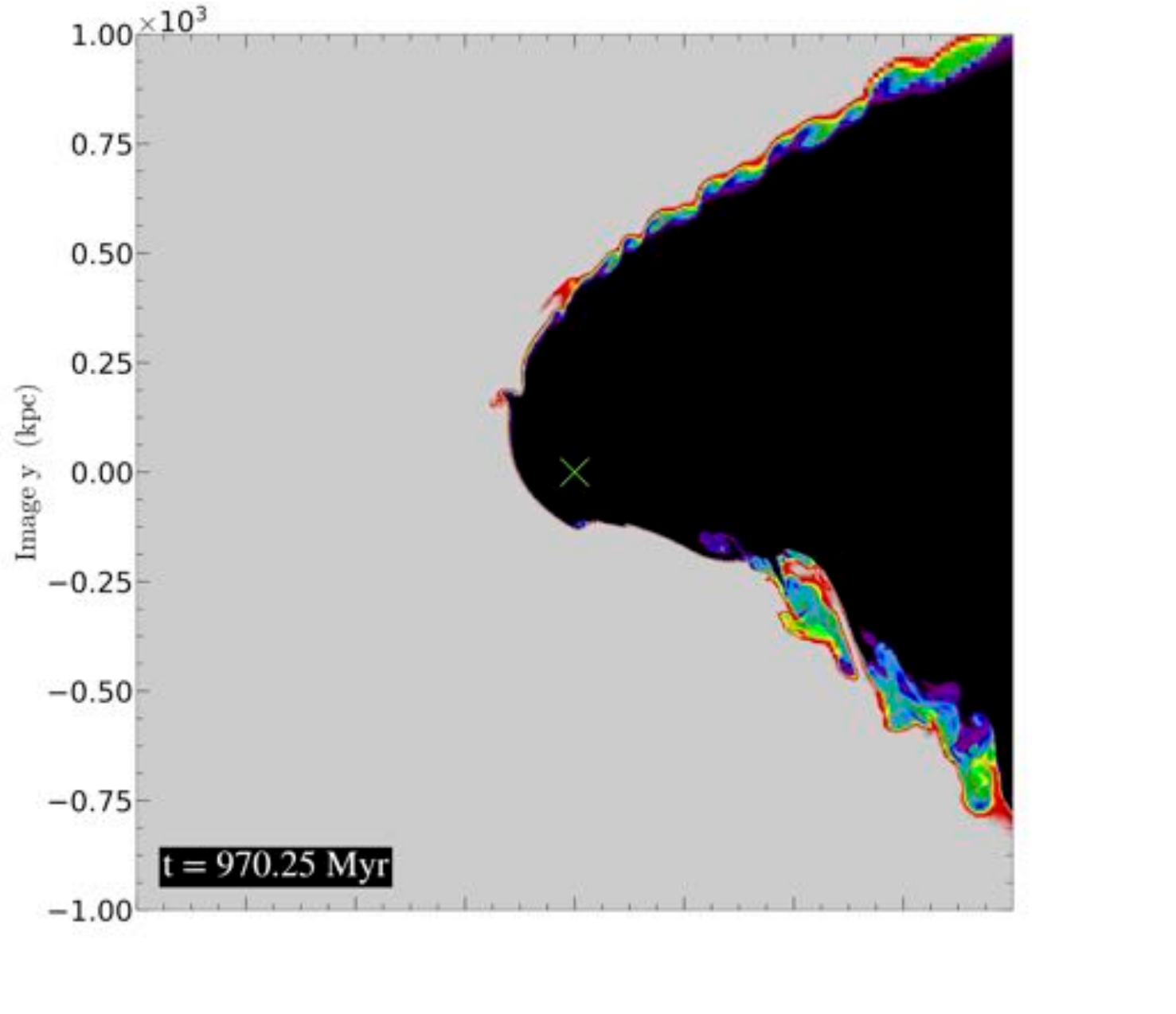}
\includegraphics[trim={0cm 1.0cm 1.7cm 0cm}, clip,width=\columnwidth]{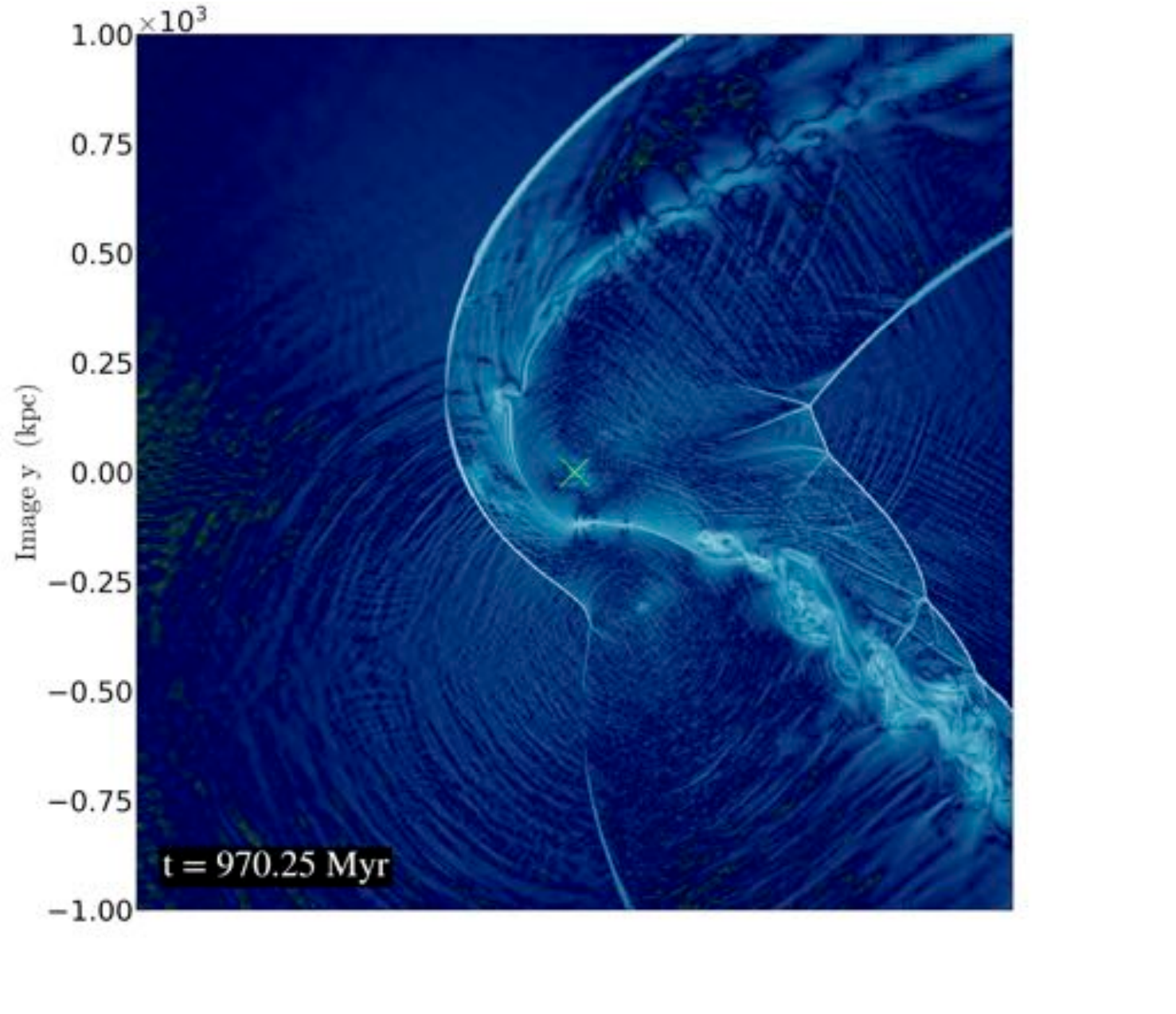}
\includegraphics[trim={0cm 0.5cm 1.7cm 0cm}, clip, width=\columnwidth]{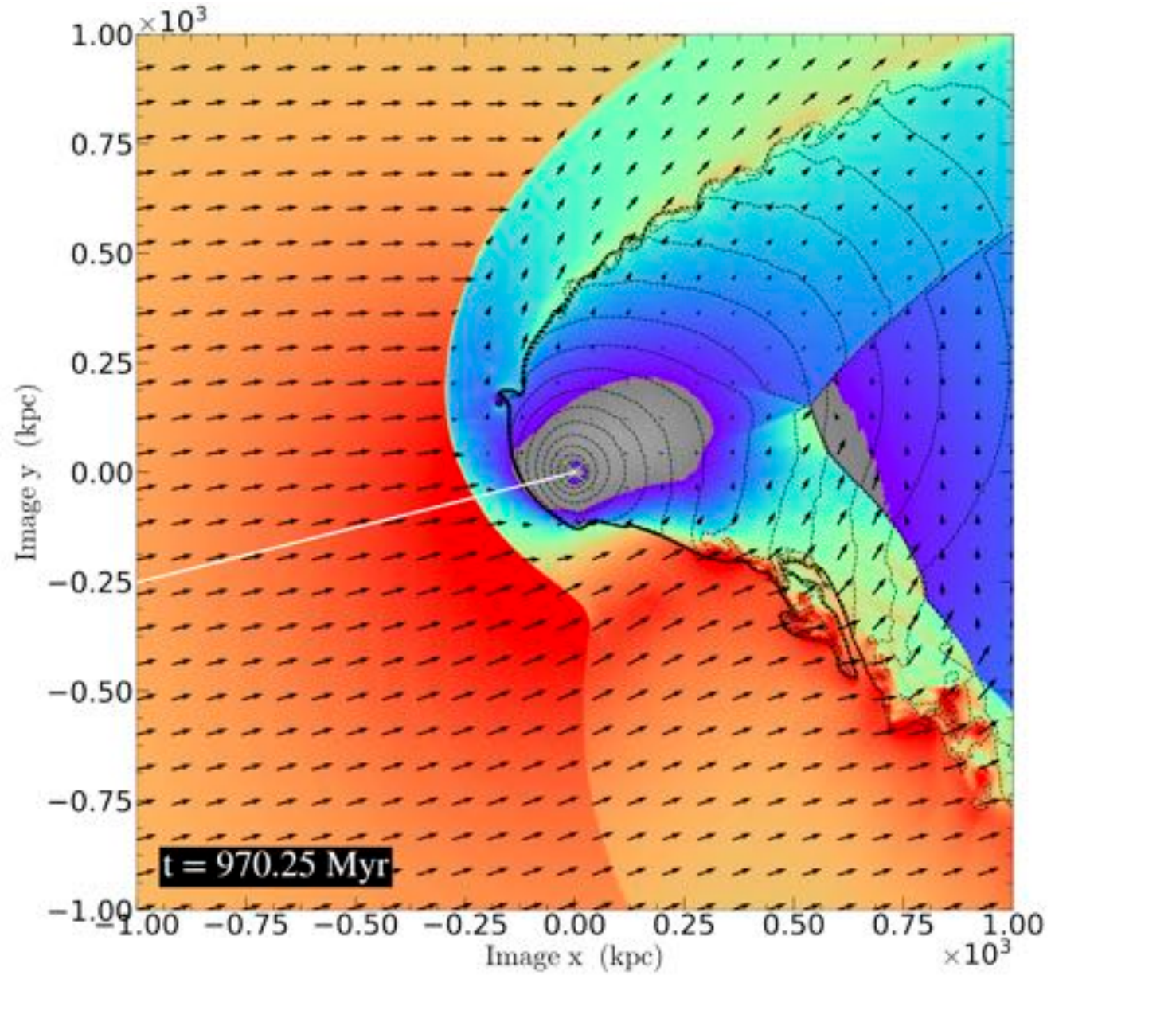}
\\(a)\\
\end{minipage}
\begin{minipage}{0.26\textwidth}
\centering
\includegraphics[trim={1.7cm 1.0cm 1.7cm 0cm}, clip, width=0.873\columnwidth]{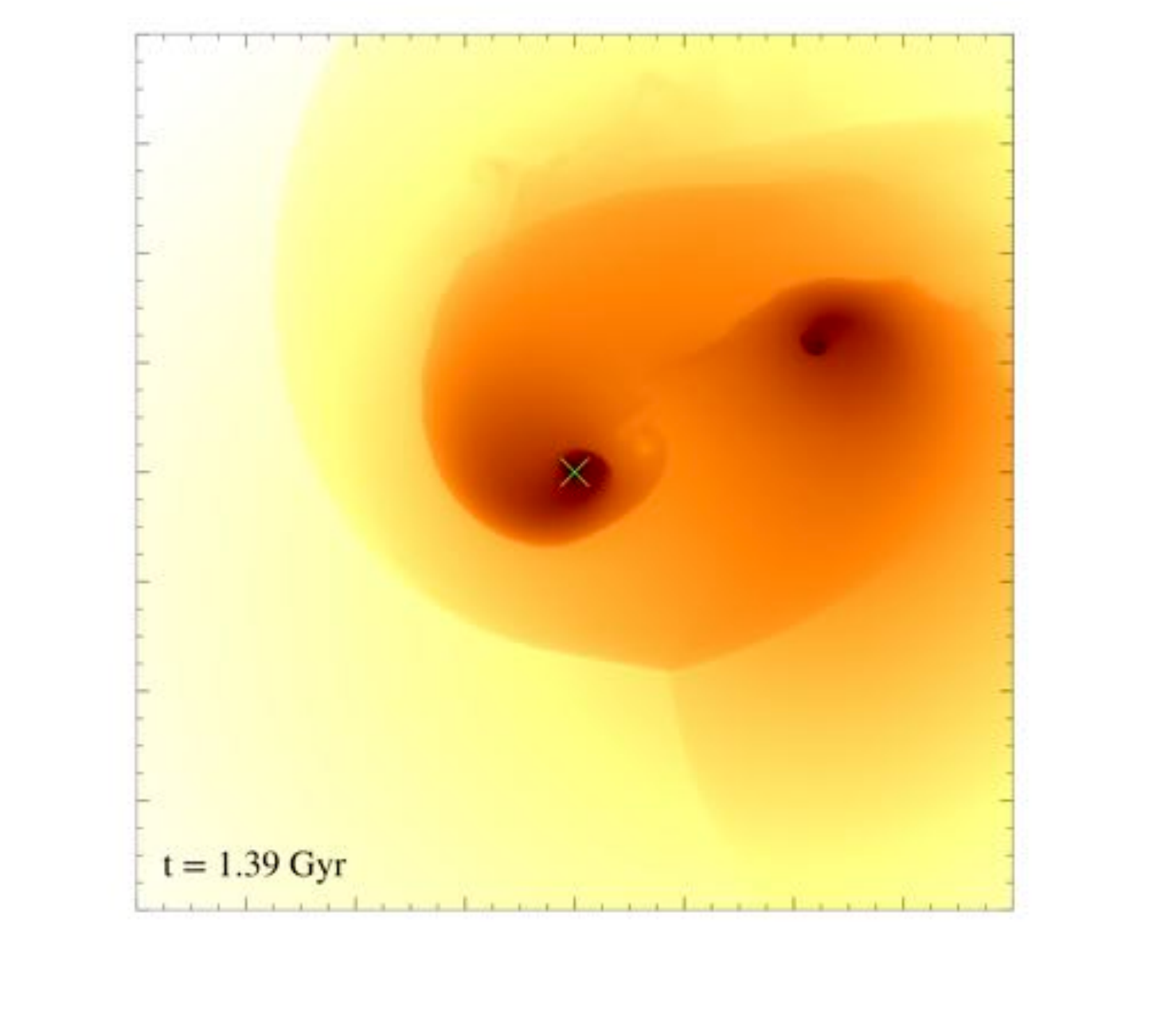}
\includegraphics[trim={1.7cm 1.0cm 1.7cm 0cm}, clip, width=0.873\columnwidth]{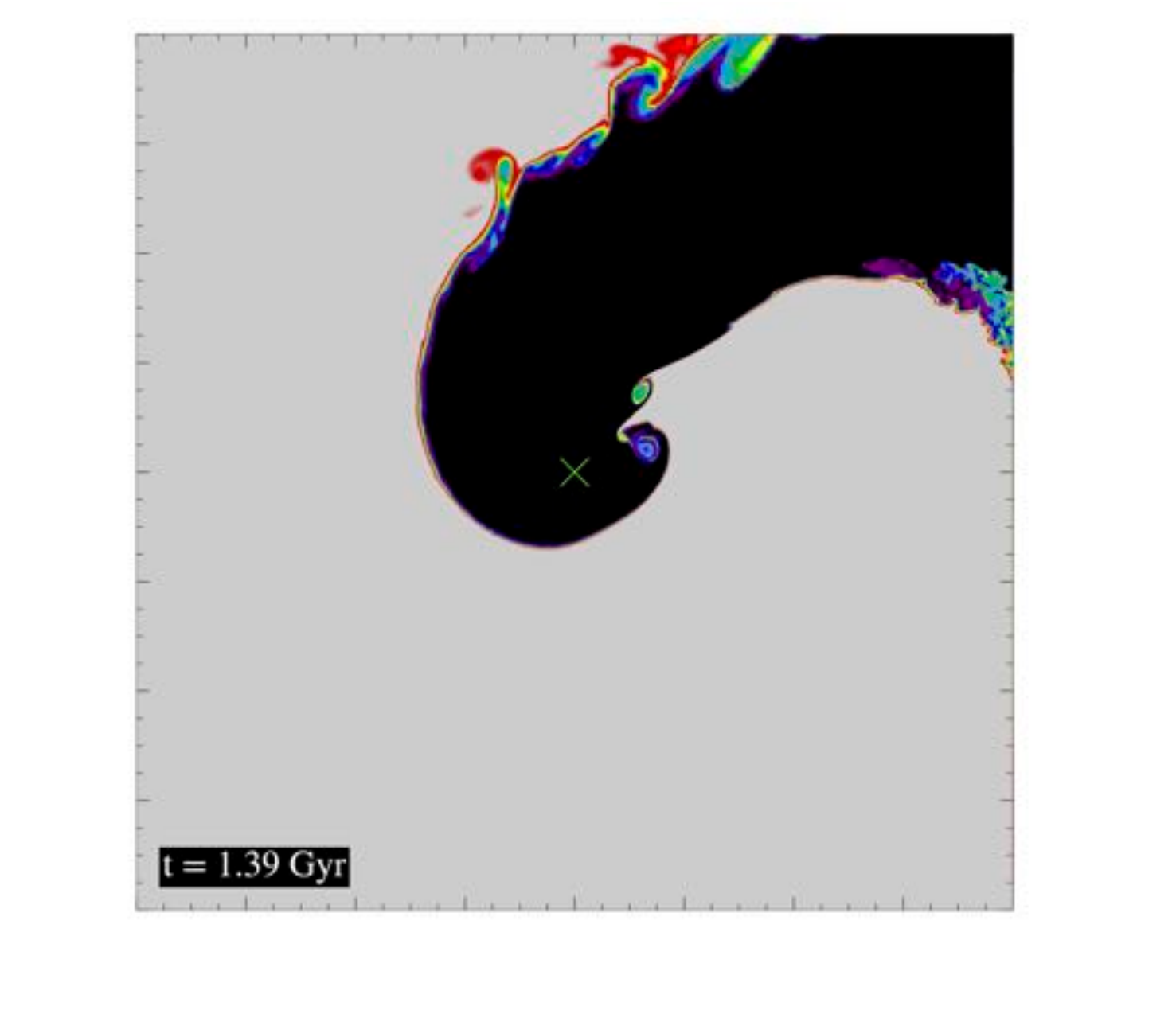}
\includegraphics[trim={1.7cm 1.0cm 1.7cm 0cm}, clip, width=0.873\columnwidth]{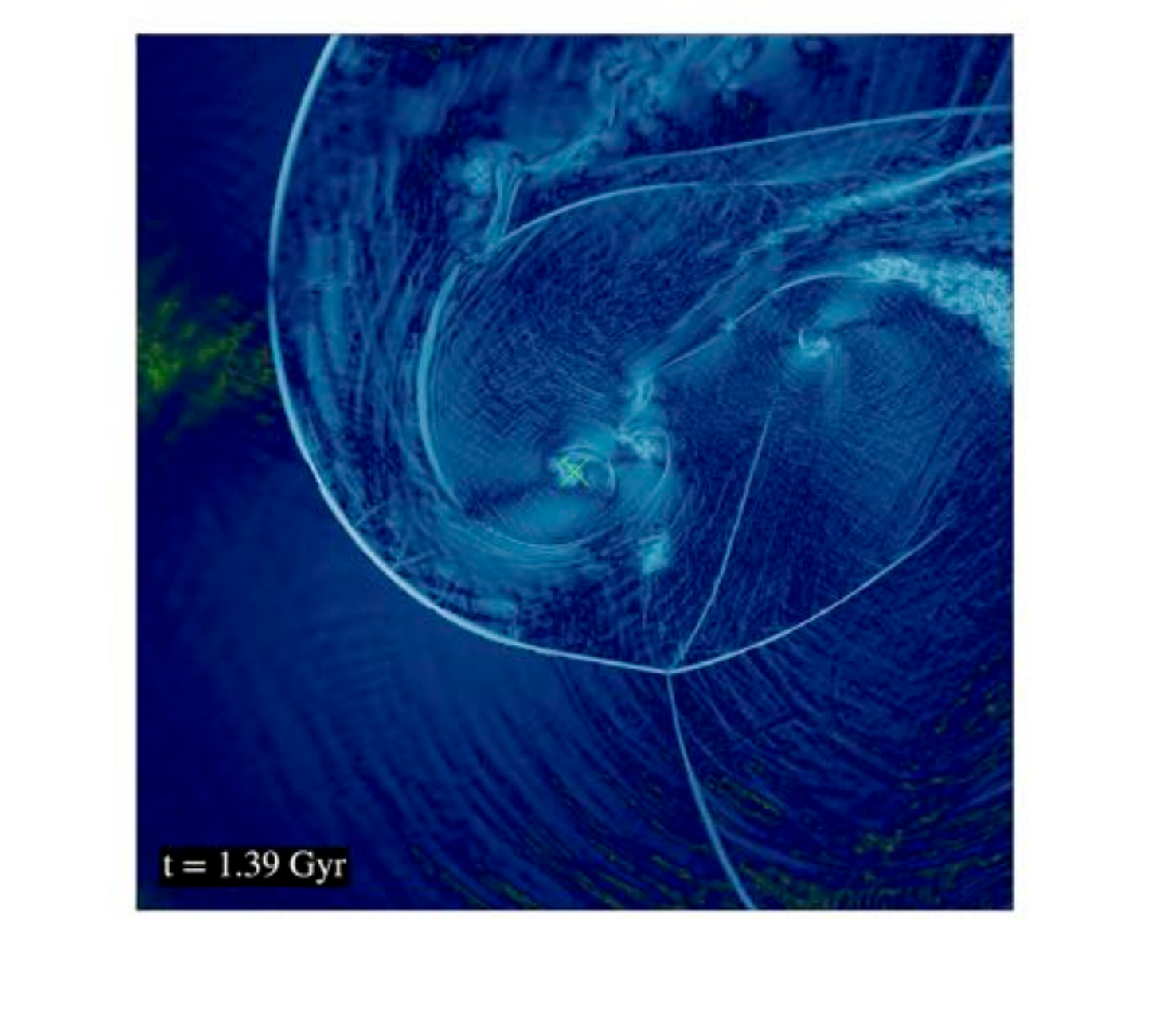}
\includegraphics[trim={1.7cm 0.5cm 1.7cm 0cm}, clip,width=0.873\columnwidth]{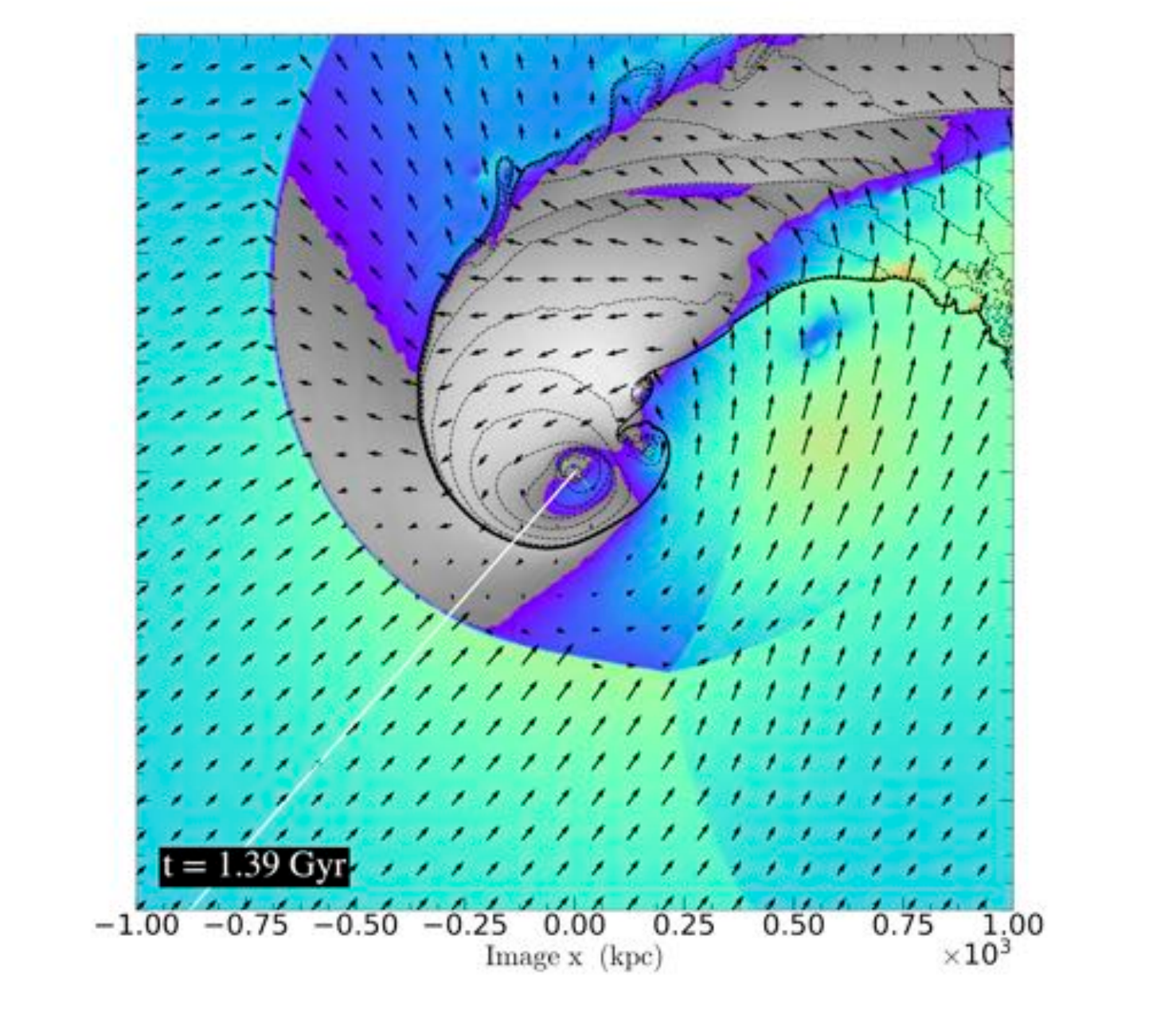}
\\(b)\\
\end{minipage}
\begin{minipage}{0.26\textwidth}
\centering
\includegraphics[trim={1.7cm 1.0cm 0cm 0cm}, clip,width=\columnwidth]{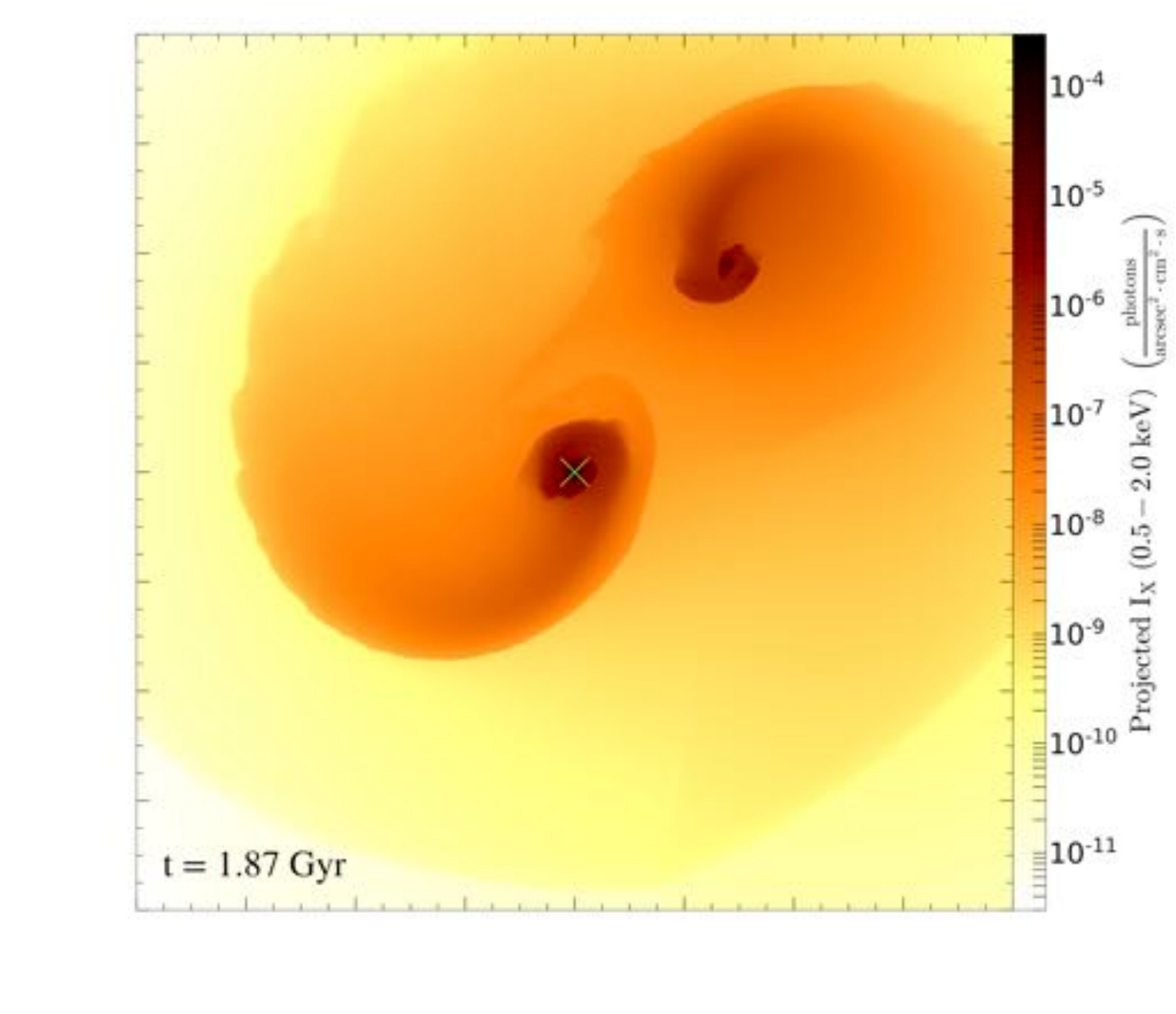}
\includegraphics[trim={1.7cm 1.0cm 0cm 0cm}, clip,width=\columnwidth]{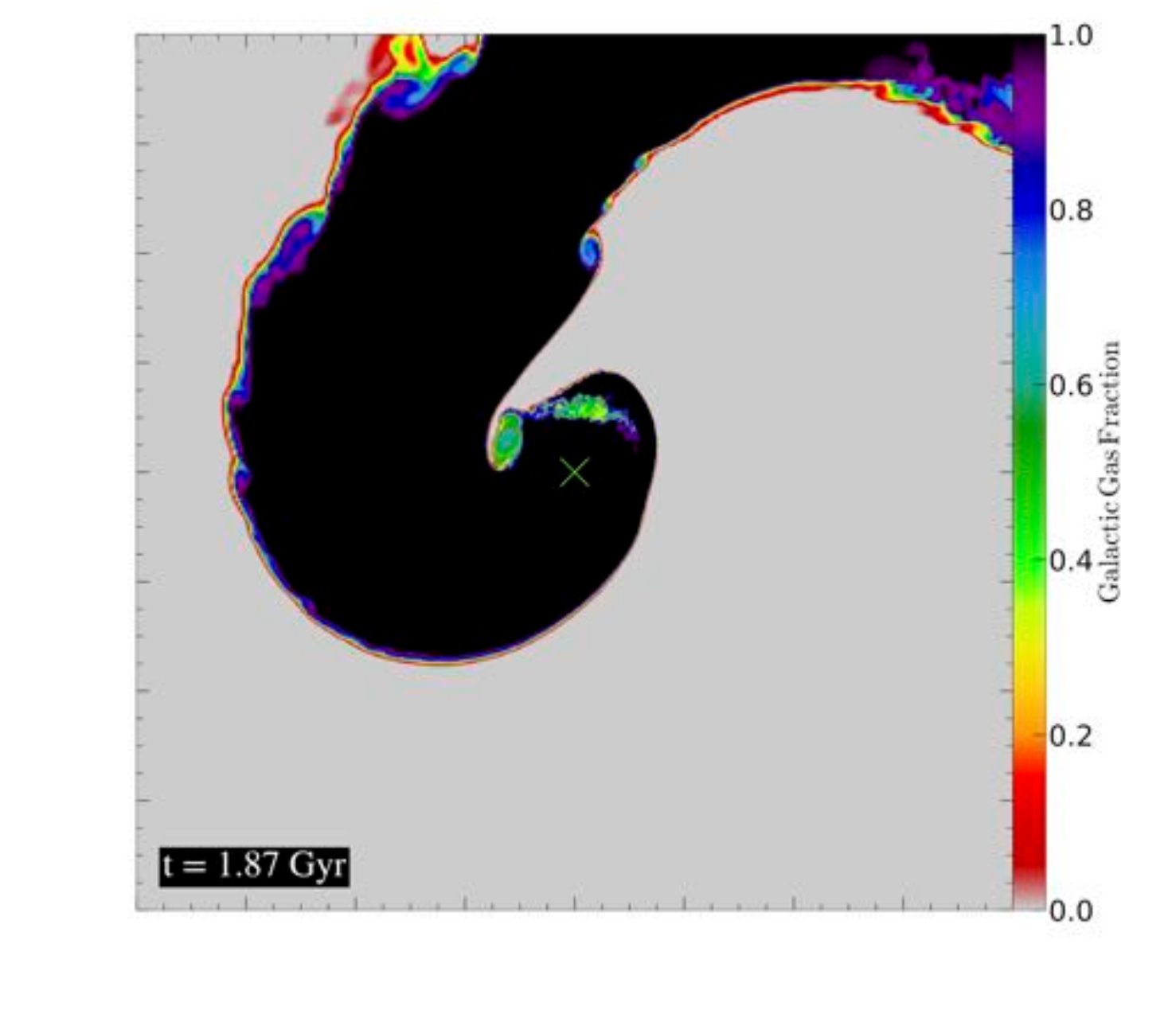}
\includegraphics[trim={1.7cm 1.0cm 0cm 0cm}, clip,width=\columnwidth]{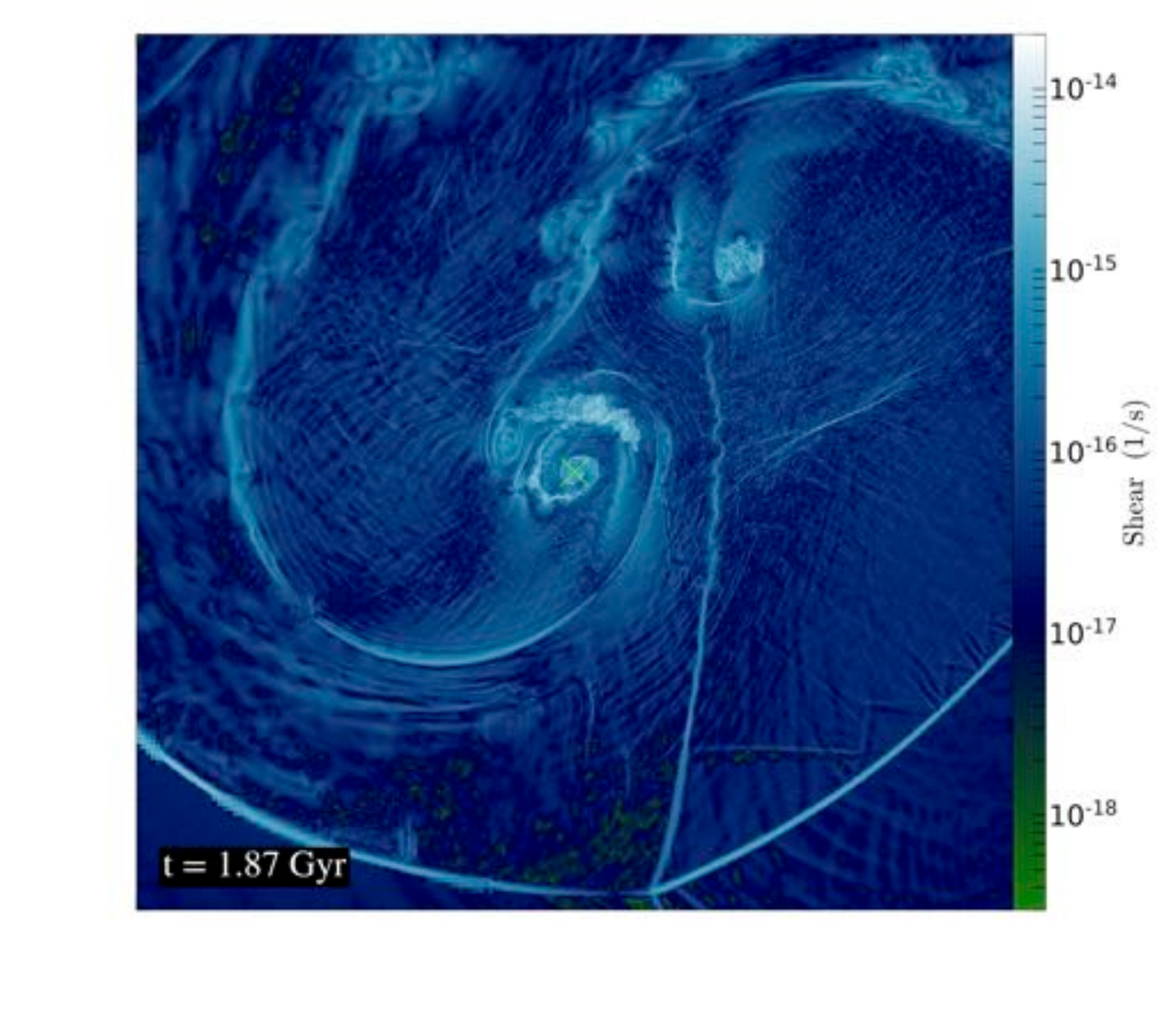}
\includegraphics[trim={1.7cm 0.5cm 0cm 0cm}, clip,width=\columnwidth]{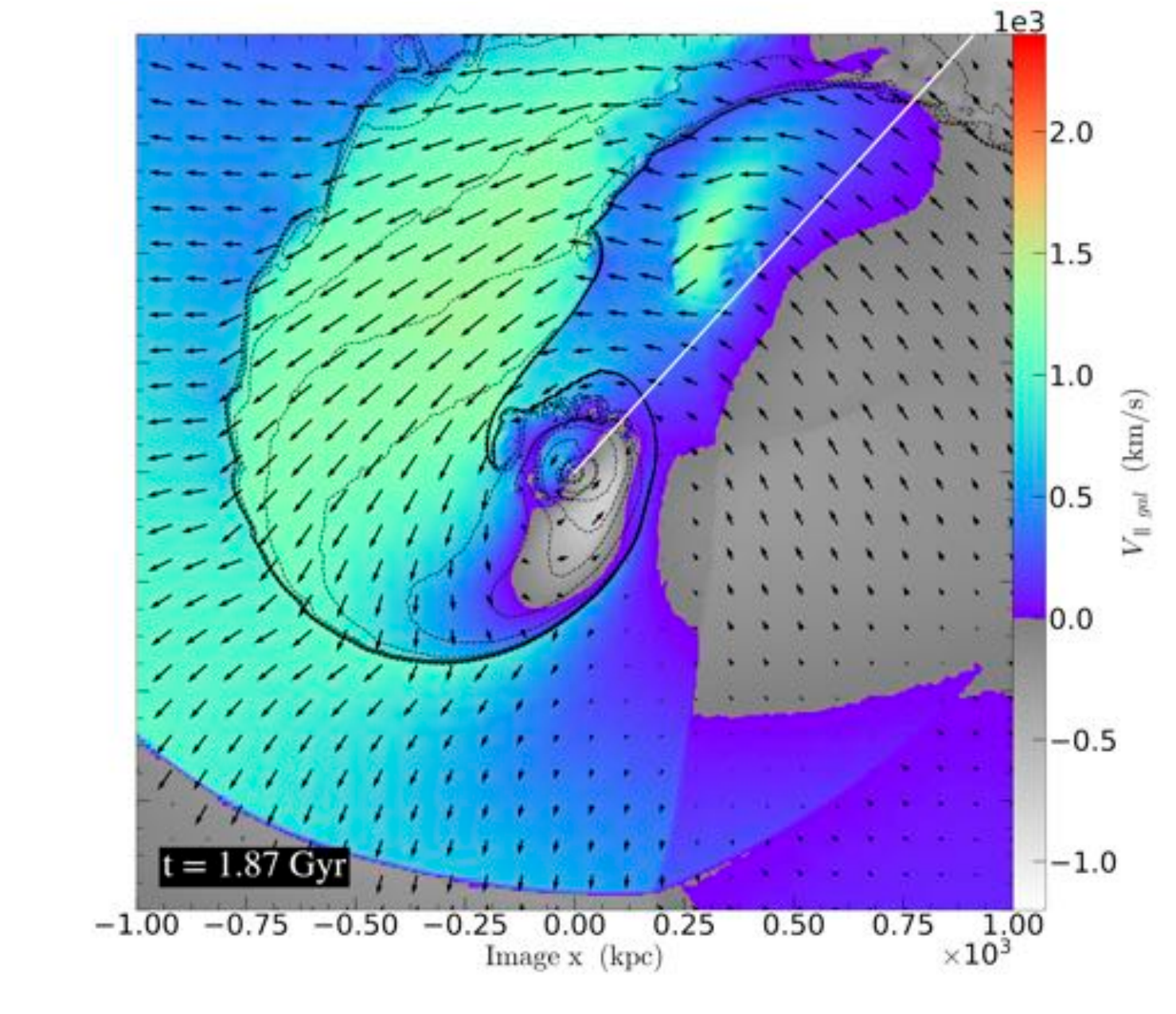}
\\(c)\\
\end{minipage}
\end{minipage}

\centering
\caption{This figure presents the \emph{arc-shaped slingshot tail} version of Figure~\ref{fig:flow}. These images are taken from the 1:3 merger simulation shown in the two right-hand columns of Figure~\ref{fig:slingshot_tails}.\\
\\
The images in Column (a) show the unstable flow beginning to develop, with a particularly asymmetric flow beyond the shock due to the location of the primary. (b) shows the secondary near apocenter, midway through the development of the slingshot tail; the secondary's tail starts to create an arc, as the outer edge of the tail is pushed out beyond the secondary. (c) shows the flow shortly before it becomes classed by this paper as a stable infall again.}
\label{fig:flow2}
\end{figure*}

Ram pressure tails or slingshot tails recently attracted interest as locations to study turbulence or its suppression in the ICM \citep{Roediger2015d, Eckert2017}. To do so, it is important to understand the principal flow conditions in and around such tails. Furthermore, the regular flow patterns around the secondary are a prerequisite to the meaningful application of the stagnation point method to determine the secondary's velocity \citep{Vikhlinin2001a, Su2017f}. In what follows, we show that this method is not applicable to secondaries that produce slingshot tails due to their complex flow patterns which differ from the classic ram pressure scenario. \par

The genesis of a slingshot tail can be split into two periods as discussed above and shown in Figures~\ref{fig:flow} and ~\ref{fig:flow2}. True for both slingshot tail forms, right after pericenter passage, the secondary continues to drag a significant amount of its downstream atmosphere along as a remnant tail. As the secondary slows and changes direction approaching apocenter, the remnant tail is carried by its momentum and its attraction to the secondary potential as it falls back toward the remnant atmosphere of the secondary. At this point, there is significant flow within the remnant tail transverse to the secondary's direction of motion, with similar flow patterns regardless of the form of slingshot tail.\par

The flow patterns in the second period are complex and potentially misleading. The secondary either develops into an \emph{arc-shaped slingshot tail}, (Figure~\ref{fig:flow2}b), or develops into an \emph{overrun slingshot tail}, (Figure~\ref{fig:flow}b), as the tail begins to fall back and wash over the secondary. For the latter, the remnant tail washes over the secondary causing a 'false' head-tail shape to form (i.e. a head-tail that does not represent the motion through the ICM); this is the process which generates the conical tail of the \emph{overrun slingshot tail}. Additionally in this process, the overrunning tail causes some stripping of the remnant atmosphere of the secondary, adding to the 'false' head-tail shape. An example of this is shown in Figure~\ref{fig:flow}b, where the sharpest edge in the X-ray plot may naively suggest a roughly north-easterly direction of motion, although the secondary moves to the south. Additionally, this process disrupts the internal structure of the secondary, as its atmosphere sloshes around its potential. Also note the complex flow patterns in the surrounding ICM which do not resemble a flow around a blunt body.

As mentioned, the beginning of the flow for both slingshot cases is similar, but there are some key differences. One such difference can be seen when comparing rows 2 and 3 in Figures~\ref{fig:flow} and ~\ref{fig:flow2}. In the \emph{arc-shaped} form, the secondary's tail holds significantly more of its own gas through pericenter passage, with an area of laminar flow following the secondary within the tail. This laminar flow appears to translate to the smooth arc-shaped edge in Figure ~\ref{fig:flow2}b,c. Conversely, the \emph{overrun} form shows a much more turbulent/broken tail (see Figure~\ref{fig:slingshot_tails}, rows 1 and 2 for a wider view of the simulation), perhaps better described as a wake at later stages, as the secondary gas is now well mixed with ICM. This is made obvious when comparing the shear rate in both slingshot forms. For the \emph{arc-shaped} form, we see that there is significantly less shear in comparison to the \emph{overrun} form as the tail gas co-moves with the ICM and the turbulent regions of the tail at the outer edge are mostly shed as the secondary reaches apocenter. We note that an \emph{arc-shaped slingshot tail} can be more turbulent if the secondary does not manage to retain such a large amount of its own atmosphere past its pericenter passage, for example Figure ~\ref{fig:hydraA}.

%While a turbulent tail is an obvious difference between the two slingshot forms discussed here, it does not necessarily lead to the tail overrunning the secondary; for example, the simulation in Figure~\ref{fig:hydraA} results in a very turbulent slingshot tail but ultimately the tail does not overrun the secondary.

%The second period flow patterns are particularly complex and potentially misleading. For example, as the tail overruns the secondary, a `false' upstream edge can form, so called as it does not represent the motion of the galaxy through the global ICM. This is the process generating the conical wake of the \emph{overrun slingshot tail}. The ram pressure stripping by the wake is misleading when determining the direction of motion and merger history. Additionally, this process disrupts the internal structure of the secondary as the secondary atmosphere sloshes around its potential.
%An example of this is shown in Figure~\ref{fig:flow}b, where the sharpest edge and near tail in the X-ray plot suggests a roughly north-easterly direction of motion, despite the flow patterns showing more significant flow from the south-east.

%The secondary then begins its next infall, back toward the primary, returning to regular, steady flow, with a textbook ram-pressure tail if it has held onto enough gas. The beginnings of the next infall can be seen in Fig.~\ref{fig:flow}c \&~\ref{fig:flow2}c.

As the secondary moves away from apocenter, starting its next infall into the primary, the flow patterns return to a quasi-steady flow state of the ram pressure stripping scenario (Figures~\ref{fig:flow}c and ~\ref{fig:flow2}c), similar to the blunt body case. Figure~\ref{fig:flow}c shows the \emph{overrun tail} during the second phase as the flow begins to return to the ram pressure stripping scenario. The flow here is fairly stable, but retains some asymmetry from the internal disruption/sloshing of the secondary and the bulk motions of the ICM; this image is chosen to illustrate the difficulty in judging whether the flow is steady.

It is worth noting that the stagnation point method \citep{ Vikhlinin2001a, Su2017f} to determine a secondary's velocity from stagnation point pressure relies on the analogy of a (quasi-)stable flow past a blunt body. The merger stage prior and near pericenter passage would qualify for this, with columns (a) in Figures~\ref{fig:flow} and ~\ref{fig:flow2} showing borderline cases. However, around apocenter the flow patterns in the ICM around the secondary are quite different and the stagnation point method is not applicable. Only when the regular flow patterns have been re-established during the next infall can the stagnation point method be applied again. Fish et al. (in preparation) discuss this point and further limitations of the stagnation point method.

\section{Distinguishing Between Ram Pressure Stripped Tails and Slingshot Tails} \label{sec:obsevational}

In this section we highlight the key observable signatures of slingshot tails, explaining how to use them to distinguish between a slingshot and ram pressure stripped tail. We remind the reader that we are concerned with the slingshot tail stage which occurs around apocenter of the orbit, i.e., we are only dealing with gas tails of secondaries that are at a large distance from the primary's cluster center, a prerequisite for identifying a slingshot tail. It is secondaries located at large distances from the primary's center that need careful consideration. \par

The main signature to distinguish between a ram pressure stripped tail and a slingshot tail is the tail orientation and morphology. As mentioned, a classic ram pressure stripped tail has an orderly head-tail structure, where the tail generally fades continuously away from the remnant atmosphere into the wake of mixed gas (see \citealt{Roediger2015c}). However, slingshot tails can point sideways or ahead relative to the direction of motion and do not fade continuously away, but rather have a sharp cut off, highlighted by the dashed line in Figure \ref{fig:slingshot_proj} showing a clear edge between the tail and the ambient ICM. If a gas tail of a secondary which is located at a large distance from the primary's center points transversely to the radius between the secondary and the cluster center, instead of radially away, a slingshot tail should be suspected. Subclusters rarely move on circular orbits with large radii which would be required to produce a ram pressure tangential to the cluster center. Such transverse tails arise naturally in the slingshot phase. Both slingshot tail forms typically have a density, temperature and brightness which is in between that of the ambient ICM and the remnant atmosphere of the secondary.\par

Another observational signature of slingshot tails is that in both slingshot tail forms, the secondary's atmosphere can show the presence of shells due to internal sloshing and re-accretion of gas. These shells are not apparent in a simple, ram pressure stripped secondary. Additionally, in the slingshot tail phase it can be difficult to identify a clear upstream edge (as described in Section \ref{sec:flow}). This is especially applicable to the overrun slingshot tail as described in Section \ref{sec:mode2}, the turbulent nature of the tail creates a phase where the secondary appears to have an irregular shaped atmosphere.\par

If present, the position of a bow shock can also help to distinguish between a ram pressure stripped or slingshot tail. In the ram pressure tail case, a bow shock leads just ahead of the secondary as it moves through the ICM of the primary cluster, shock heating the gas. For a slingshot gas tail, this is not the case. As the secondary slows toward apocenter, the previously leading bow shock continues to propagate outwards as the secondary turns around, hence the shock appears behind the secondary on the tail side, not leading it, and can be found at large distances behind the secondary (the shock can be located up to 1 Mpc behind the secondary). Such a detached bow shock is visually marked in Figure \ref{fig:slingshot_tails} and in the simulations of \citet{Sheardown2018}.\par

If the secondary has a slingshot tail, the primary's cluster center should show signs of the earlier core passage of the secondary. For large mass ratios, this could be the onset of sloshing and for low mass ratios, the primary could form a slingshot tail of its own. After the first passage, the cluster core will show elongation in the direction toward the secondary. If the pericenter passage was close enough, this may have even destroyed the central core. If the secondary has completed a second passage of the cluster center, the sloshing in the core will have evolved further, producing a prominent cold front on the opposite side of the cluster to the secondary. Further, the wake of the secondary could appear as a characteristic brightness edge in the primary, marking roughly the secondary's orbit (see \citet{Sheardown2018}). However, a caveat to using the dynamical state of the cluster to help identify a slingshot tail is that it would only work with a simple cluster setup, i.e. an ideal case of a binary merger or few possible merger candidates. For a system which has many merger partners it would be too difficult to attribute features of the cluster core to one single candidate.\par

\section{Classifying Some Known X-ray Tails} 
\label{sec:stails}
In this section, we argue that some examples of gas tails reported in previous papers are likely slingshot tails instead of classic ram pressure stripped tails. We note that at this stage, our arguments and comparisons are purely qualitative. A full confirmation of our suggestions may require tailored simulations to reproduce the observations quantitatively.

\subsection{LEDA 87745 in Hydra A}
\begin{figure}
\centering
\includegraphics[scale=0.32]{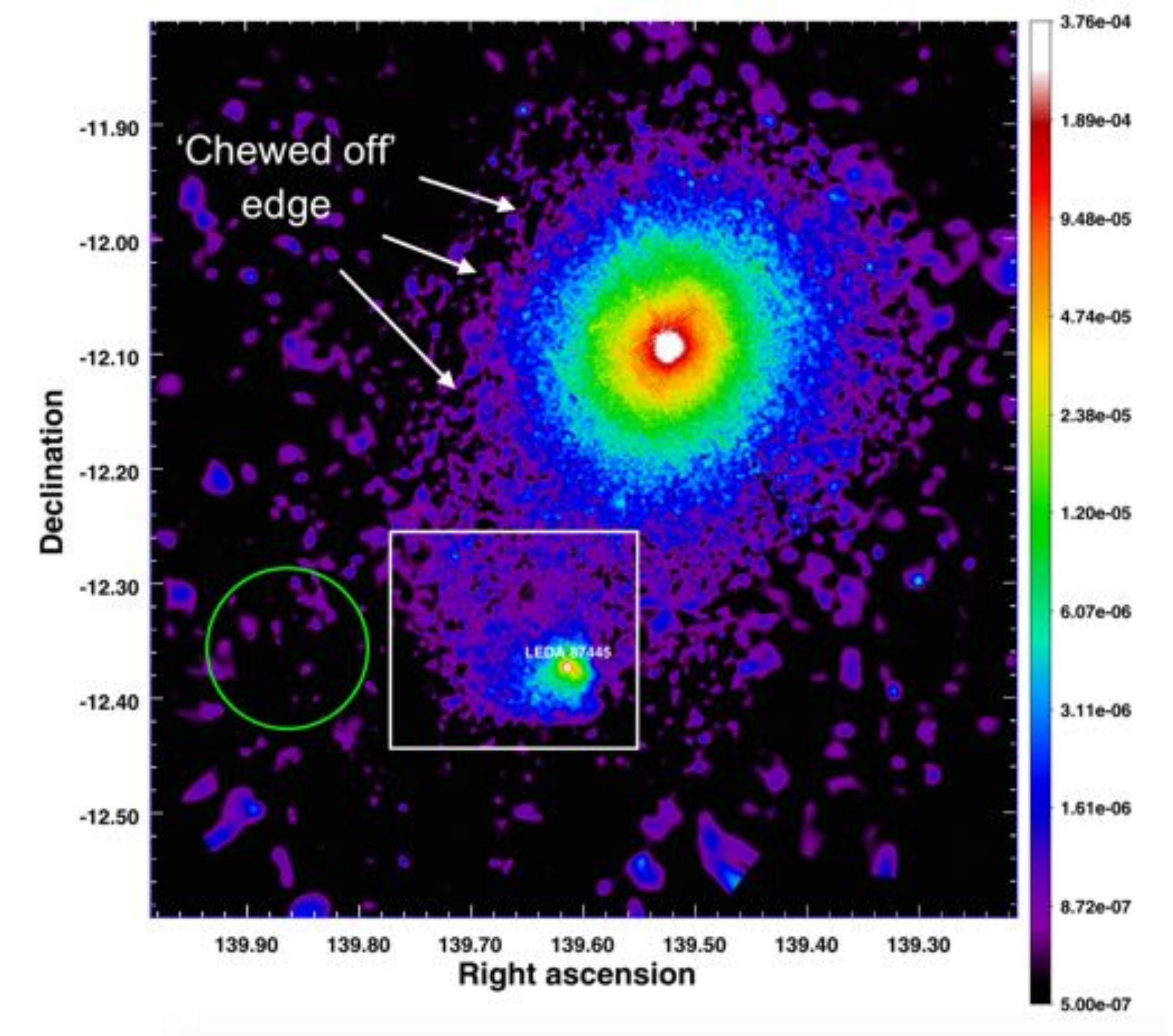}
\includegraphics[scale=0.32]{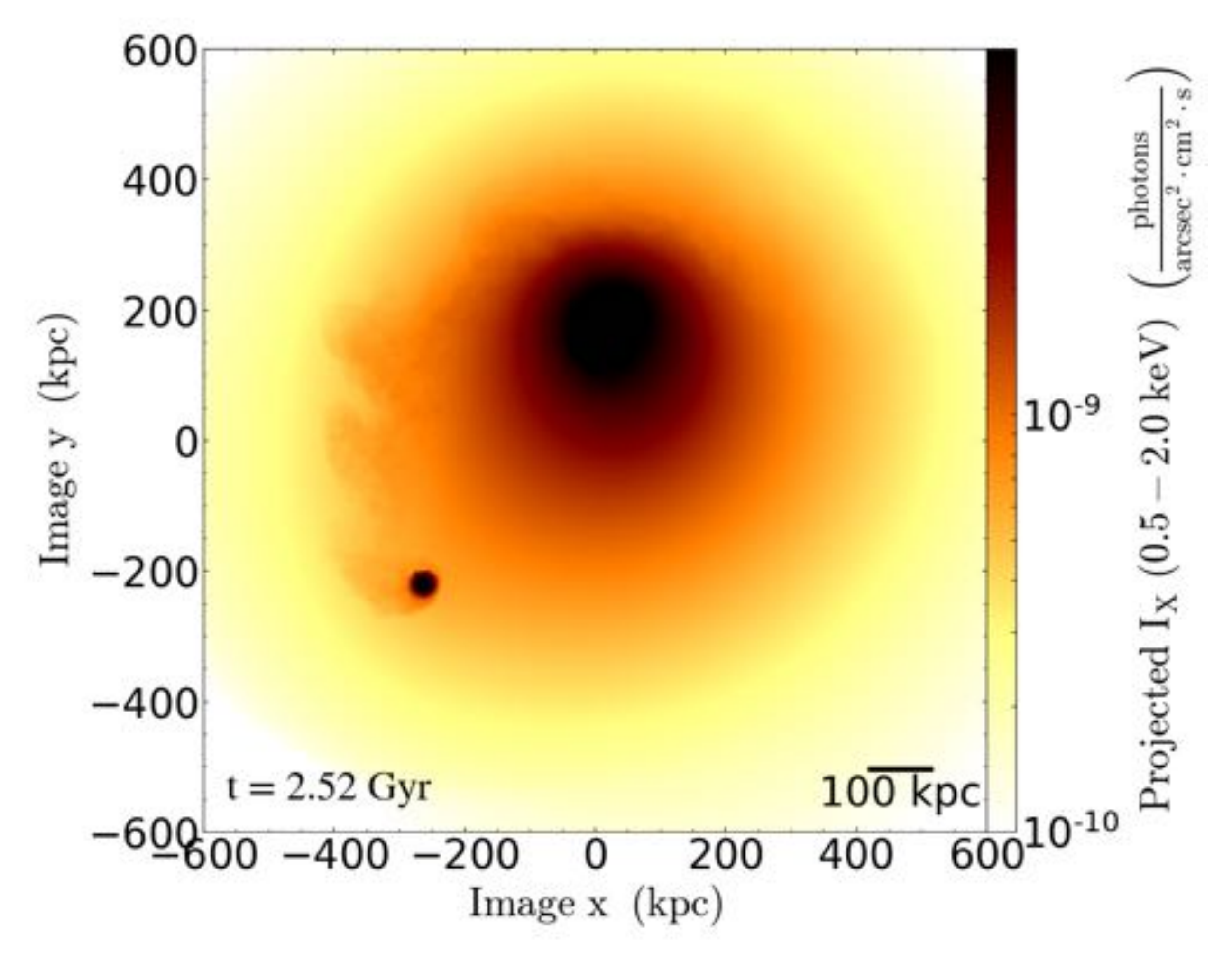}
\caption{Top: Image taken from \citet{DeGrandi2016b}. An adaptively-smoothed, vignetting-corrected XMM/EPIC mosaic image of the Hydra A Cluster in the 0.7-1.2 keV band. Bottom: X-ray photon intensity projection made from the V2 simulation in \citet{Sheardown2018}. This simulation image is chosen to provide a visual match to the observational features of the tail in LEDA 87445 as the secondary reaches apocenter. The cluster in the simulation also shows elongation of the primary towards the secondary, much like the image of the Hydra A Cluster.}
\label{fig:hydraA}
\end{figure} 

Located 1.1 Mpc south of the Hydra A Cluster center, LEDA 87445 is the dominant member of a galaxy group with a gas tail about 760 kpc long (\citealt{DeGrandi2016b}) that demonstrates several features which resemble a slingshot tail in action. Firstly, the galaxy group is at a large distance from the cluster center, and from Figure \ref{fig:hydraA}, we see the tail direction is transverse to the radius joining LEDA 87445 to the cluster center. If this tail direction is taken to indicate the direction of motion, the transverse orbit would be hard to explain. Further evidence for a slingshot tail is the dynamical state of the cluster. An offset central AGN shock (\citealt{Nulsen2005} and \citealt{Simionescu2009}) toward the north of the cluster indicates large scale bulk motions and an east-west asymmetry is apparent showing a 'chewed off' edge in the east as indicated in Figure \ref{fig:hydraA}. The observed asymmetry implies LEDA 87745 passed by the cluster center from the north-east with a large impact parameter which created the 'chewed off' edge, and as the galaxy group moved out to the apocenter, it produced the observed slingshot tail. In Figure \ref{fig:hydraA}, we provide a visual simulation match to LEDA 87745 using the V2 simulation in \citet{Sheardown2018}. The secondary in this simulation has a turbulent arc-shaped tail because due to its low mass, it could not retain a very large remnant tail past pericenter passage. As mentioned in Section \ref{sec:mode1}, this could be an example of an intermediate case which lies inbetween the arc-shape and overrun slingshot forms. If the slingshot tail scenario is correct for this case, there should be a detached bow shock located south LEDA 87445 (in the direction away from the cluster center) at a distance of $>$ 750 kpc.

\subsection{NGC 7618 and UGC 12491}
\begin{figure}
\centering
\includegraphics[scale=0.35]{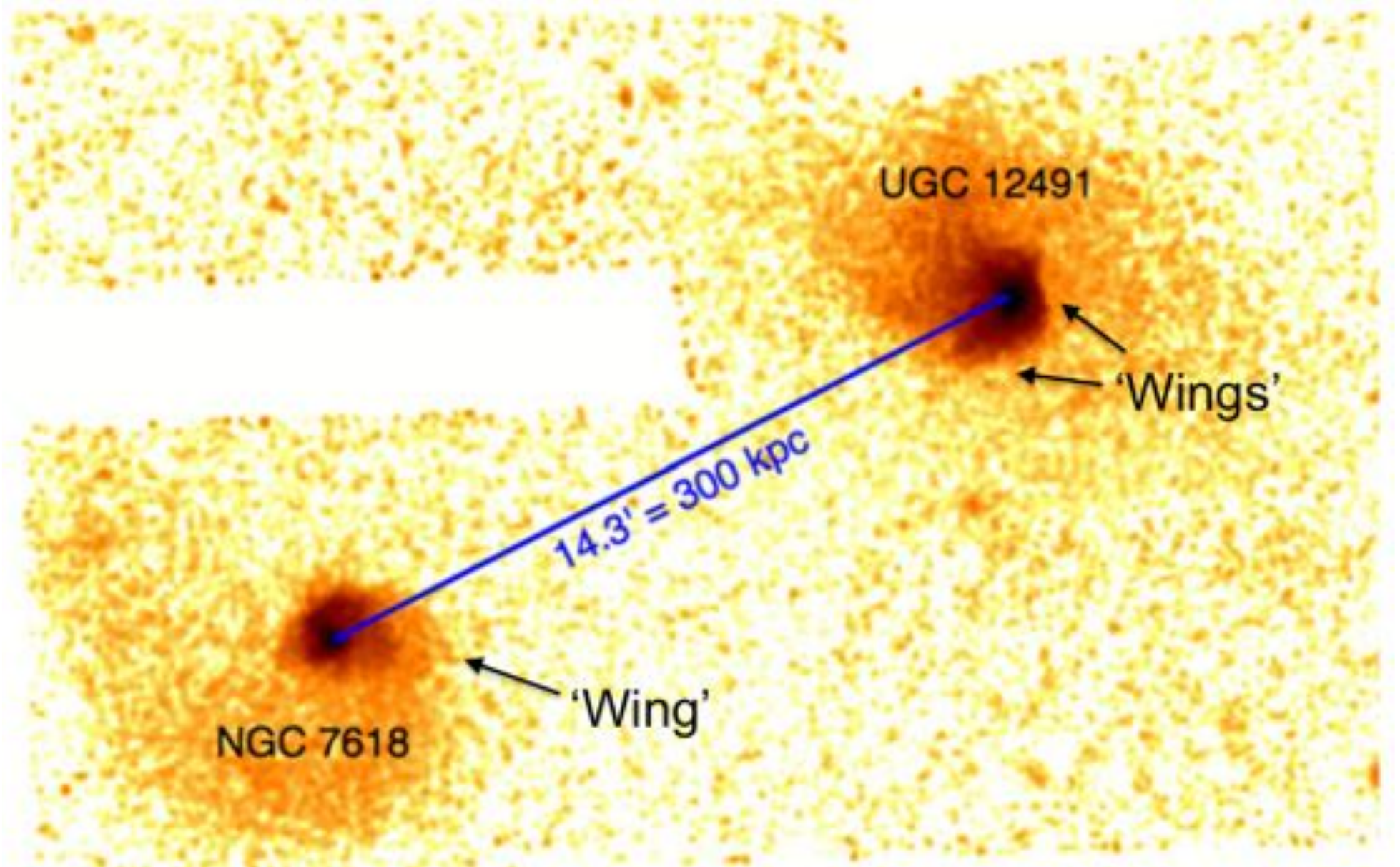}
\includegraphics[scale=0.30]{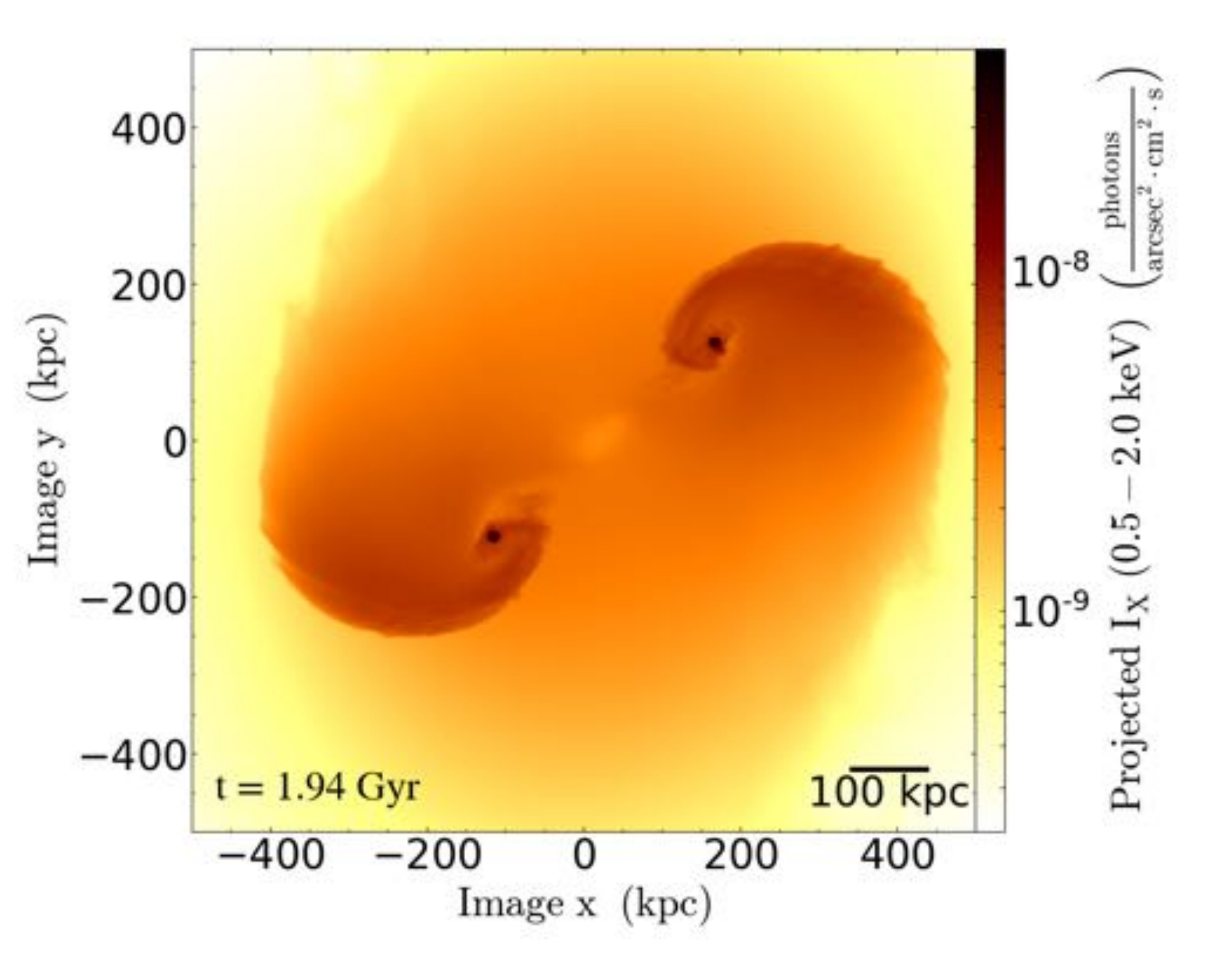}
\includegraphics[scale=0.30]{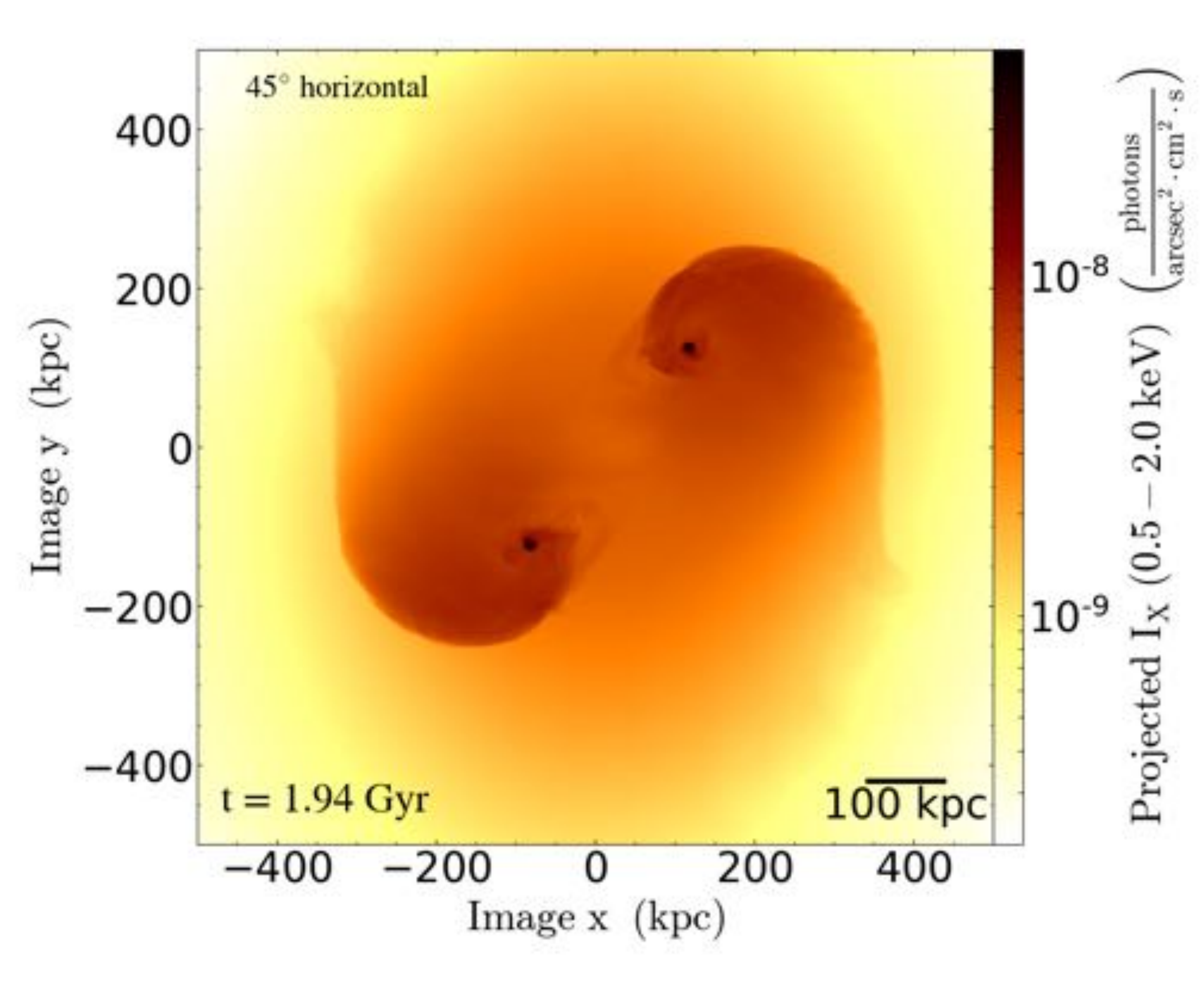}
\caption{Top: Image taken from \citet{Roediger2012}. A co-added, background-subtracted, and exposure-corrected 30 ks Chandra/ACIS-S image of the NGC 7618 and UGC 12491 galaxy groups in the 0.5-2.0 keV band, smoothed with a 4 arcsec Gaussian kernel. Middle: A simulated X-ray photon intensity projection for a 1:1 merger with a pericenter distance of 265 kpc between two clusters with a mass of 6$\times$10$^{13}$M$_{\odot}$ as setup in \citet{Sheardown2018}. Bottom: Likewise but the simulation image is rotated by 45$^{\circ}$. The simulation images shows the two clusters at the first apocenter stage of their merger and reveals prominent arc-shaped slingshot tails in both, providing a visual match to the observed image. Rotating the merger plane by an angle of 45$^{\circ}$ accounts for the more highly wound arc-shaped tails observed here. Given that the merger partners are at apocenter, no relative velocity between the two is expected, as observed in NGC7618 and UGC12491.}
\label{fig:ngc7618}
\end{figure} 

Shown in Figure \ref{fig:ngc7618}, NGC 7618 and UGC 12491 are at the centers of merging galaxy groups of approximately equal mass. Using Chandra observations, \citet{Roediger2012} found that the pair both displayed arc-like sloshing cold fronts and $\sim$ 100 kpc long spiral tails. The authors also suggest that since the cores of both groups are not destroyed, that the encounter between them was not a close one. From our analysis, we find that arc-shaped slingshot tails are produced only when the impact parameter is large, as is likely the case here. With these ideas in mind, we ran a 1:1 merger simulation with a large impact parameter (a pericenter distance of 265 kpc) using the cluster setup as in \citet{Sheardown2018}. This cluster setup was chosen simply for its roughly similar mass to NGC 7618 and UGC 12491.

An X-ray projected image from the simulation is shown in Figure \ref{fig:ngc7618} and we can see that it provides an excellent match to the observed features as it clearly replicates the arc-shaped tails and position of the groups. Therefore, we propose that these are not simple sloshing cold fronts, but rather arc-shaped slingshot tails and both groups are at apocenter of their merging orbit. Based on their original idea of sloshing cold fronts, \citet{Roediger2012} therefore suggested that there should be Kelvin-Helmholtz instabilities (KHI) along the spiral tails of both groups. However, as shown in our simulation, with the arc-shaped slingshot tail form, there are only a few regions with strong shear, and KHIs form only slowly near the far end of the arced tails. Therefore, using a different pointing of Chandra to look further down the spiral tail could perhaps reveal the presence of KHI in this system. As shear flows along the apparent cold fronts or slingshot tails may appear in different locations, it is important to distinguish between both scenarios for studying the presence of KHI or their suppression by ICM microphysics, such as viscosity or draped magnetic fields.

\subsection{NGC 4839 Group in Coma}

\begin{figure}
\centering
\includegraphics[scale=0.38]{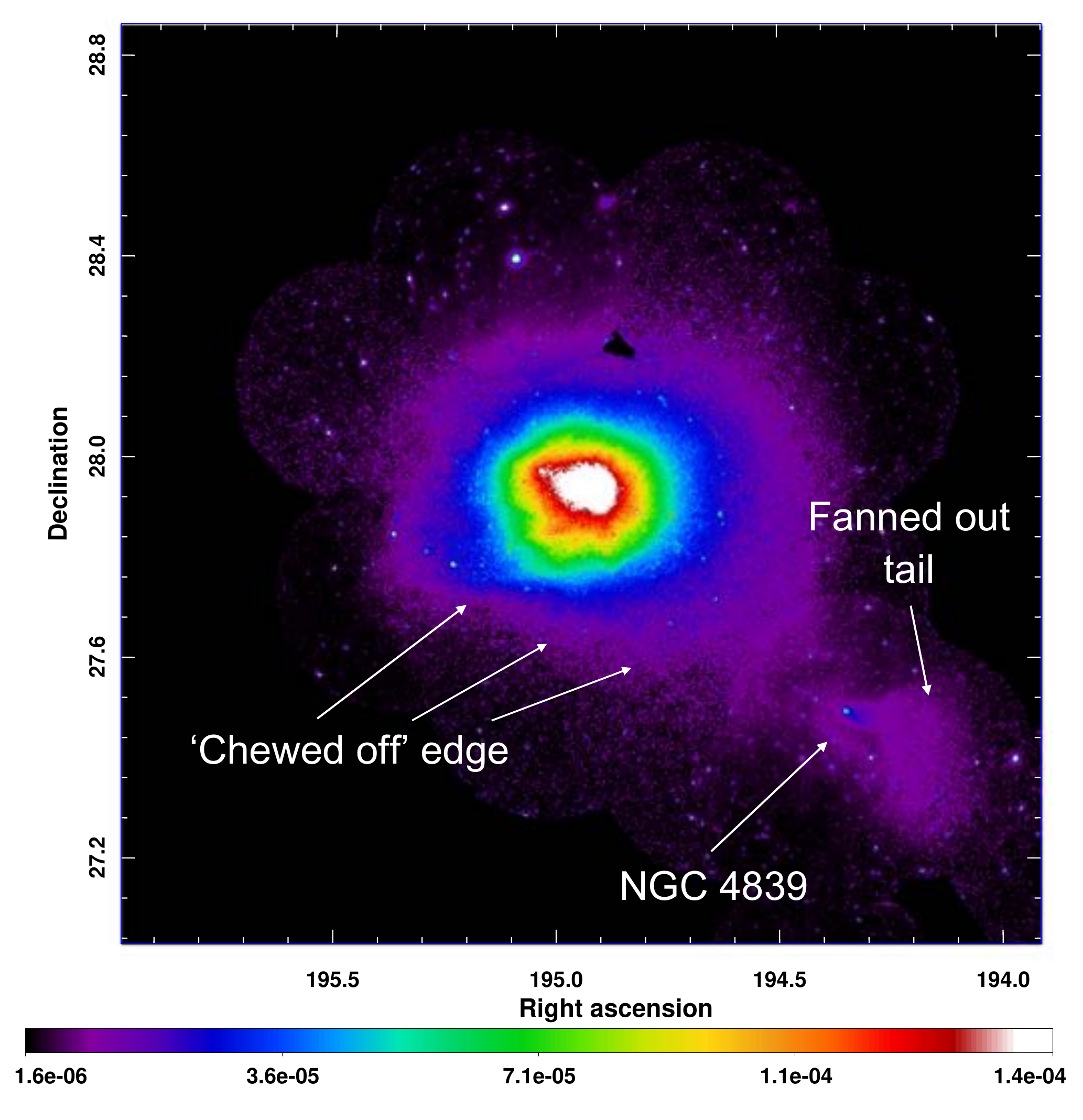}
\includegraphics[scale=0.34]{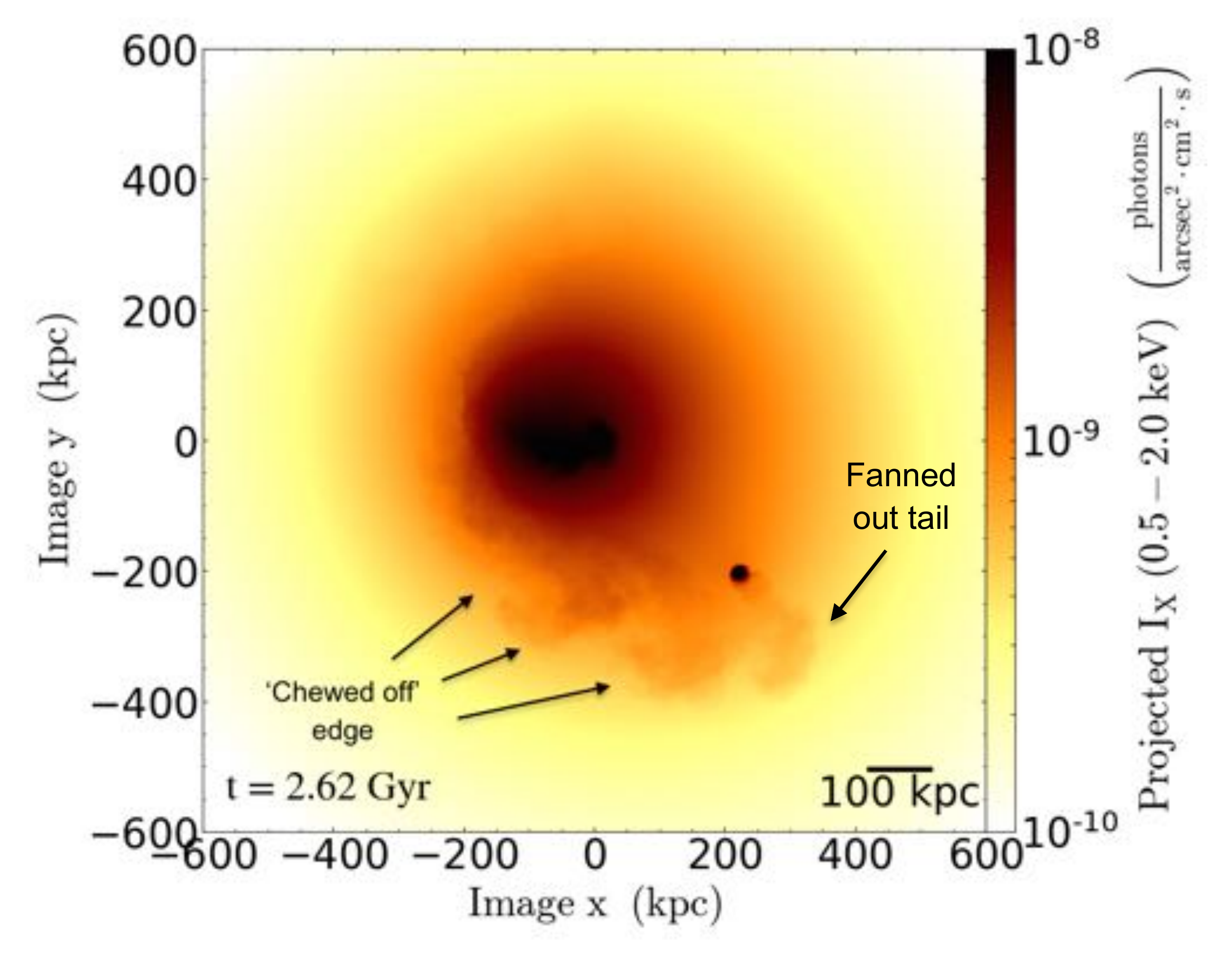}
\caption{Top: Image taken from Lyskova et al. (submitted). An XMM-Newton image of the Coma Cluster and the NGC 4839 group in the 0.5-2.5 keV energy. Bottom: X-ray photon intensity projection from the V1 simulation in \citet{Sheardown2018}. The simulated image shows the overrun slingshot tail in action and we see that the tail geometry matches the observed tail of NGC 4839. Physical scales are different as this simulation is not tailored to the Coma Cluster. See Lyskova et al. (submitted) for the tailored simulation of Coma and the NGC 4839 group.}
\label{fig:ngc4839}
\end{figure}

In the outskirts of the Coma Cluster lies the galaxy group NGC 4839, approximately 1 Mpc in projection south-west from the cluster center (\citealt{Neumann2003}, Lyskova et al. submitted). As shown in the top image of Figure \ref{fig:ngc4839}, the group is merging with the cluster as X-ray images reveal a truncated atmosphere along with a $\sim$ 600 kpc elongated tail of cool gas which is homogeneous in brightness and temperature and fanned out in the direction away from the group \citep{Sasaki2016}.

Thus we have several features for this case which resemble an overrun slingshot tail instead of a ram pressure stripped tail which was previously thought. Furthermore, the far edge of this fanned out tail marks the maximum radius the tail has slingshotted to. The truncated atmosphere would suggest the group has fallen through the cluster once already. Additionally, a radio relic was discovered near the virial radius of the Coma cluster, 2 Mpc in projection from the cluster center, far beyond NGC 4839, but in the same south-west direction as the group \citep{Akamatsu2013}. This radio relic could potentially be the detached bow shock of the galaxy group. Therefore, we propose that the group passed by the cluster center from the east with a small impact parameter, went into the overrun slingshot tail form and is now on its next infall. In Figure \ref{fig:ngc4839}, we show a simulated X-ray projection from the V1 simulation in \citet{Sheardown2018} in this slingshot stage to provide a visual match to NGC 4839. The idea that the tail of NGC 4839 is not due to ram pressure, but due to the group falling through the cluster once has also been confirmed independently by Lyskova et al. (submitted).

\begin{figure}
\centering
\includegraphics[scale=0.37]{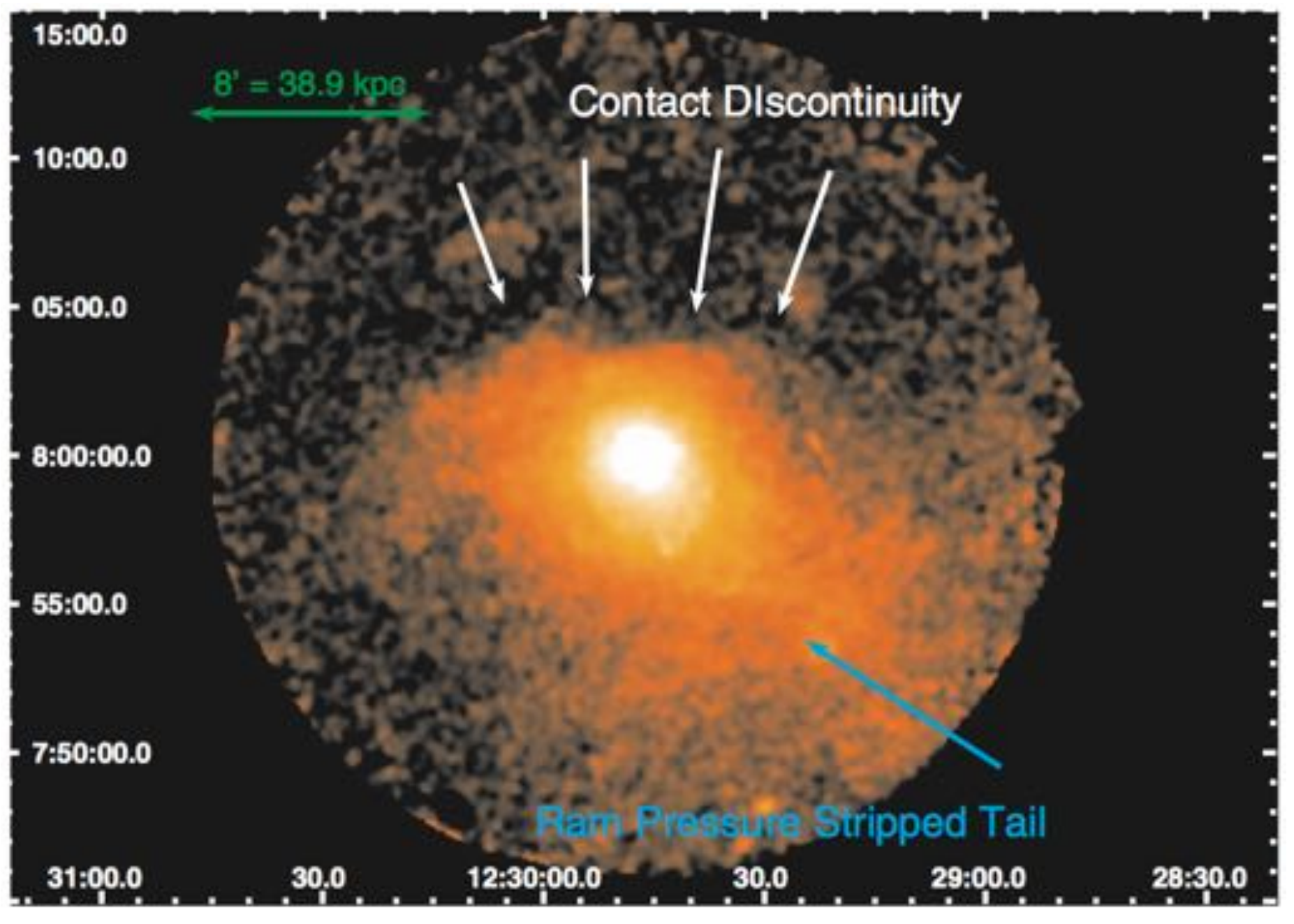}
\includegraphics[scale=0.42]{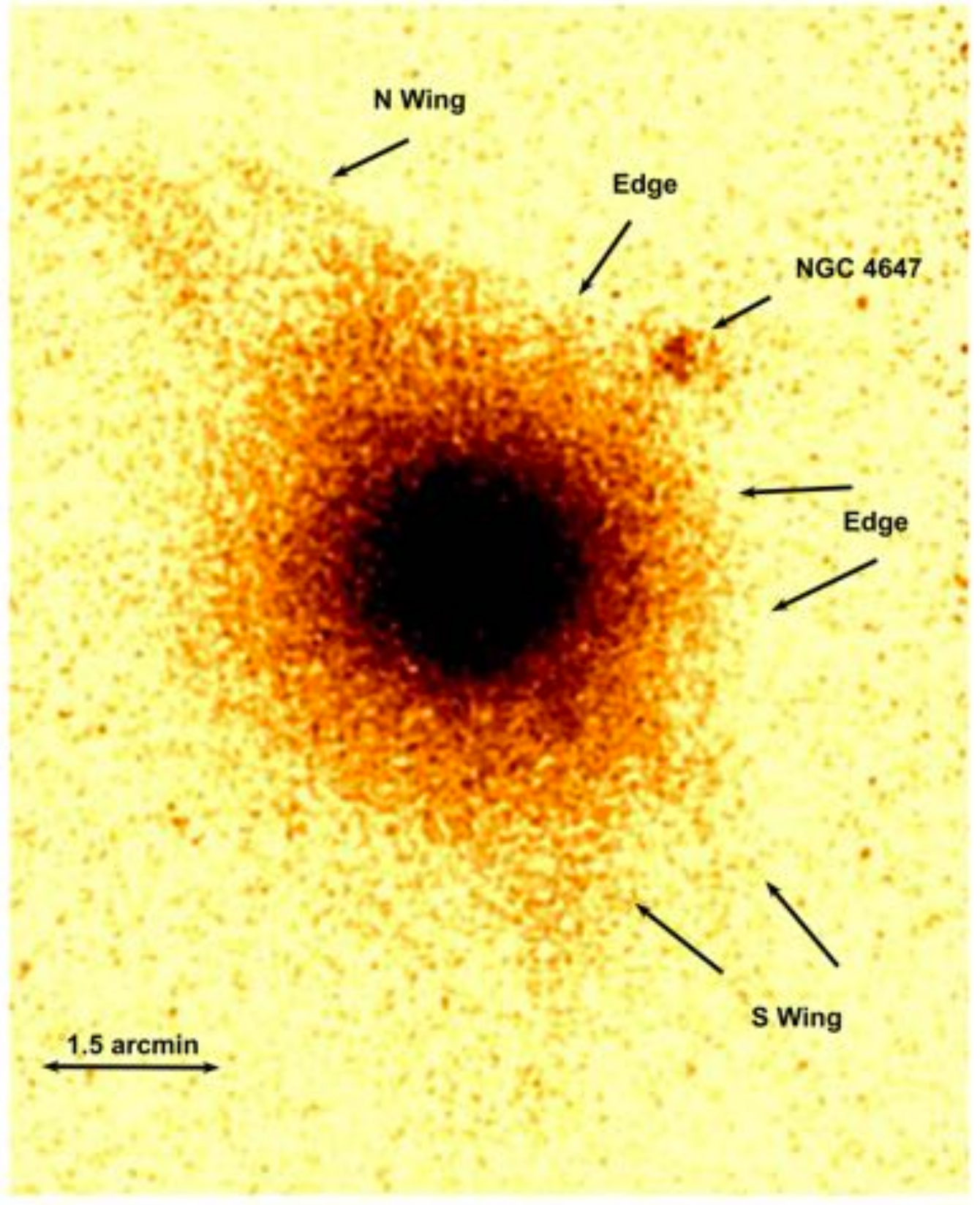}
\caption{Top: Image taken from \citet{Kraft2011a}. An exposure corrected, Gaussian smoothed XMM-Newton image of M49 in the 0.5-2.0 keV band with point sources removed. Bottom: Image taken from \citet{Wood2017}. An Exposure-corrected, background-subtracted, coadded Chandra X-ray image of M60 in the soft band (0.5-2.0 keV).}
\label{fig:virgo}
\end{figure}

\subsection{NGC 4472/M49 and NGC 4649/M60 in Virgo}

For these two early-type galaxies it is unclear whether or not they do indeed have slingshot tails, here we only offer a possible suggestion that a slingshot tail scenario can be applied. Shown in the top image of Figure \ref{fig:virgo}, M49 lies $\sim$ 1 Mpc south of the Virgo Cluster center and has a $\geq$60 kpc long tail pointing somewhat transversely to the radius between M49 and the cluster center, which has been attributed to ram pressure stripping (\citealt{Kraft2011a}). In the bottom image of Figure \ref{fig:virgo}, M60 is located $\sim$ 1 Mpc to the east of M87, the cluster center, and shows a truncated atmosphere and no gas tail. The evidence for a slingshot scenario is that M49 and M60 are located at large distances from the Virgo cluster center and have clearly truncated atmospheres which would be unusual for a first infall. M49 appears to have a tail which points transversely to the radius between it and the cluster center which could be an arc-shaped slingshot tail, although the tail does not appear to be a prominent arc as we have shown for this form, so this seems unlikely. Perhaps this could be a turbulent arc-shaped case as with LEDA 87745, i.e. M49 was strongly stripped on its first passage. M60 arguably has a fuzzy atmosphere, like the first phase of the overrun slingshot tail. For both cases, it could be that they are at a less favorable viewing angle, or on their third infall into the cluster, as this would give a truncated atmosphere with little or no gas tail, but given the distance to the cluster center this would be unlikely.

\begin{figure}
\centering
\includegraphics[scale=0.52]{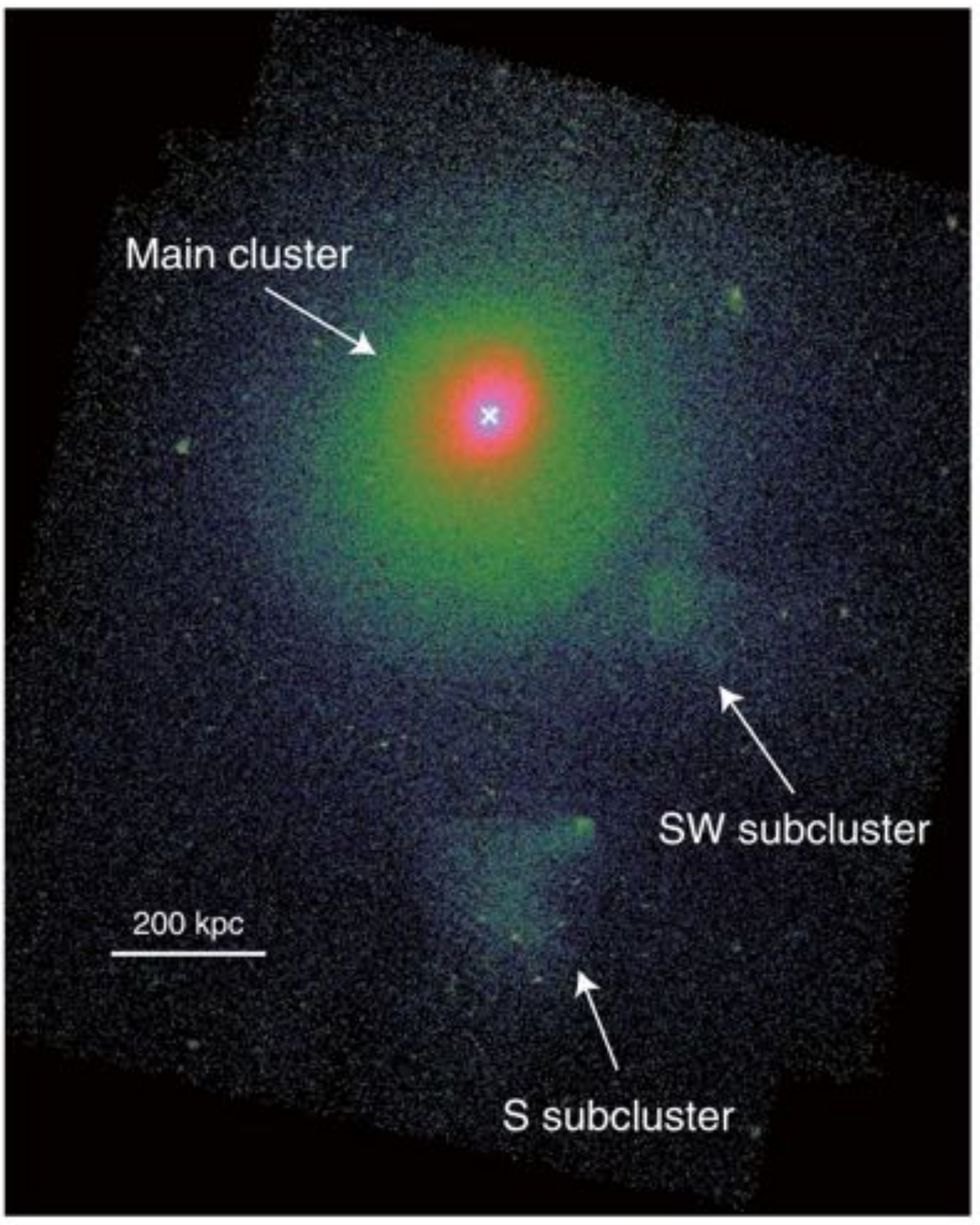}
\caption{Image taken from \citet{Ichinohe2015a}. A Gaussian-smoothed, exposure and vignetting-corrected, background-subtracted Chandra image of Abell 85 in the 0.6-7.5 keV energy band.}
\label{fig:abell85}
\end{figure}

\subsection{Abell 85}
A dynamically evolving, cool core cluster, Abell 85 boasts an array of merger features, substructures and filaments ( \citealt{Yu2016}). Figure \ref{fig:abell85} shows two prominent merging subclusters located to the south and south-west of the cluster center and also the elongation of the cluster core in the direction of the subcluster to the south. Analysing Chandra, XMM-Newton and Suzaku observations of the cluster, \citet{Ichinohe2015a} find ongoing sloshing in the cluster core which spirals out to 600 kpc that was likely triggered by merger events which occured several Gyr's ago. The authors find that the subcluster to the south is $\sim$ 600 kpc in projection from the cluster center, moving close to the plane of the sky and has a clear X-ray tail pointing to the south-east, perpendicular to the cluster center. They further determine that the outer gas of the subcluster has already been stripped away and now it is the low-entropy core that is being stripped. The stripped gas forms a gas tail which is $>$ 200 kpc in length which appears to be fanned out in the downstream direction and has an abrupt drop in surface brightness at the end of the tail. \citet{Ichinohe2015a} analysis of the tail determined that the tail has been bent and pushed eastwards due to the velocity field of the ongoing sloshing in the cluster. Thus, taking all of these features into account, it indicates a possible slingshot tail in action. The first indication is that the subcluster has been stripped of its outer gas already, suggesting it has already passed through the cluster once. This idea could be supported by the cluster's elongation towards the south, in the direction of the subcluster. Second, the tail has a fanned out shape that has an abrupt drop in surface brightness at the end of the tail, which would correspond to an overrun slingshot tail. Although the orientation of the tail perpendicular to the cluster core would not coincide with this overrun slingshot scenario as we would expect the overrun tail to be found south of the subcluster in correlation to its northward motion. However, as mentioned, \citet{Ichinohe2015a} indicate that sloshing has bent and pushed the tail eastwards into its observed position to the east/south-east of the subcluster, therefore the tail could well have been located south of the subcluster, fitting the overrun slingshot tail scenario. If this were the case, the subcluster will have an begun its merger with the cluster from the north, passing by the cluster center on its eastern side, before reaching its current southern position. This scenario could well be similar to that of the NGC 4839 group in Coma.

\section{Conclusion} \label{sec:conclusion}
In this paper we have visually inspected a suite of idealized binary cluster merger simulations to show that as well as ram pressure stripped tails, there is a second class of gas tails, named slingshot tails. These tails are formed as a secondary subhalo moves \emph{away} from the primary cluster center, toward the apocenter of its orbit, producing tails which can at times point perpendicular or opposite to the subhalo's current direction of motion. Importantly, whilst in the slingshot tail stage, the secondary is not being subjected to ram pressure stripping and the morphology of the tail is influenced more by tidal forces than ram pressure. In consequence a tail observed in the slingshot tail stage should not be identified as a gas stripping tail as this does not give an accurate description of the ongoing physics. We find that slingshot tails differ from ram pressure tails in the following way.

\begin{itemize}
\item Ram pressure stripped tails have an orderly head-tail morphology in contrast to slingshot tails which are generally oriented radially but can well point transverse to the radius between the secondary and the primary cluster center while the secondary is at a large distance from the primary cluster center.

\item The brightness of slingshot tails has a distinct end, unlike ram pressure stripped tails which continuously fade away.

\item For a ram pressure stripped tail, a bow shock will lead the secondary, whereas, for a slingshot tail, the shock that once led the secondary continues to propagate outward as the secondary turns around and heads back toward the cluster, therefore the shock appears behind the secondary on the tail side and can be found at large distances.

\item The remnant atmosphere of secondaries with slingshot tails can show evidence of shells in the remaining gas core due to internal sloshing and re-accretion of gas.

\end{itemize} 

From our analysis, we find that slingshot tails can be split into two characteristically different forms:

\begin{itemize}
\item Arc-Shaped: This form occurs when the impact parameter is large and produces a prominent arc-shaped tail which can temporarily point perpendicular to the secondary's motion (as shown in Figure 3, the top image of column (b) in the arc-shaped tail section).

\item Overrun: This form occurs when the impact parameter is small and can be separated into two distinct phases. The first phase produces an irregular shaped secondary atmosphere, as the slingshot tail overruns the remnant core of the secondary and partially settles into its potential. The second phase is reached as the remnant tail continues to overrun the core of the secondary, becoming conical in shape, fanning outward along the orbit direction, away from the secondary. The edge of the fanned out tail marks the cut off radius which the secondary has overshot to.
\end{itemize}
Furthermore, we find that in the slingshot tail stage, flow patterns around the subhalo are highly irregular. Thus, interpreting an observed slingshot tail using a simple ram pressure stripped tail scenario leads to incorrect conclusions regarding subhalo velocity or expected locations of shear flows, instabilities or mixing. Future work will involve a deeper investigation to derive the exact conditions as to why one or the other slingshot form occurs, or any other intermediate regime for that matter. Understanding slingshot tails can provide an insight into the gas physics at the cluster outskirts and also help disentangle the merger history of galaxy clusters. 

%Therefore, with the new X-ray instruments coming online in the next decade such as XARM and Athena XIFU, further work will involve making images of slingshot tails using mock instruments such as these to probe the level of understanding into these areas that can be achieved.

\acknowledgments

We thank the anonymous referee for helpful comments to improve the clarity of the paper. \\

The software used in this work was in part developed by the DOE NNSA-ASC OASCR Flash Center at the University of Chicago. \software{FLASH (Fryxell et al. 2000, 2010)} \\

E.R. acknowledges the support of STFC, through the University of Hull's Consolidated Grant ST/R000840/1 and access to viper, the University of Hull High Performance Computing Facility.\\

N.L. acknowledges partial support by grant No. 18-12-00520 from the Russian Scientific Foundation. \\

A.S. thanks Hazel Morrish for providing the cartoon image in Figure 1 to distinguish between the main slingshot tail forms. \\

\bibliography{mendeley}



\end{document}